\RequirePackage{ifpdf}
\ifpdf 
\documentclass[pdftex]{sigma}
\else
\documentclass{sigma}
\fi

\usepackage{float}

\begin{document}

\numberwithin{equation}{section}

\allowdisplaybreaks

\renewcommand{\PaperNumber}{070}

\FirstPageHeading

\renewcommand{\thefootnote}{$\star$}

\ShortArticleName{Hamilton--Jacobi Theory and Moving Frames}

\ArticleName{Hamilton--Jacobi Theory and Moving Frames\footnote{This paper is a
contribution to the Vadim Kuznetsov Memorial Issue `Integrable
Systems and Related Topics'. The full collection is available at
\href{http://www.emis.de/journals/SIGMA/kuznetsov.html}{http://www.emis.de/journals/SIGMA/kuznetsov.html}}}

\Author{Joshua D. MACARTHUR~$^\dag$, Raymond G. MCLENAGHAN~$^\ddag$ and Roman G. SMIRNOV~$^\dag$}

\AuthorNameForHeading{J.D. MacArthur, R.G. McLenaghan and R.G. Smirnov}

\Address{$^\dag$~Department of Mathematics and Statistics, Dalhousie University,\\
$\phantom{^\dag}$~Halifax, Nova
Scotia, Canada, B3H 3J5}
\EmailD{\href{mailto:joshm@mathstat.dal.ca}{joshm@mathstat.dal.ca}, \href{mailto:smirnov@mathstat.dal.ca}{smirnov@mathstat.dal.ca}}
\URLaddressD{\url{http://www.mathstat.dal.ca/~smirnov/}}

\Address{$^\ddag$~Department of Applied Mathematics, University of Waterloo,\\
$\phantom{^\ddag}$~Waterloo, Ontario, Canada, N2L 3G1}
\EmailD{\href{mailto:rgmclena@uwaterloo.ca}{rgmclena@uwaterloo.ca}}

\ArticleDates{Received February 03, 2007, in f\/inal form May
14, 2007; Published online May 24, 2007}

 \Abstract{The interplay between the Hamilton--Jacobi theory of orthogonal separation of variables
 and the theory of group actions is investigated based on concrete examples.}


\Classification{37J35; 53C12}

\begin{flushright}
\it In memory of Vadim Kuznetsov
\end{flushright}

\renewcommand{\thefootnote}{\arabic{footnote}}
\setcounter{footnote}{0}

\section{Introduction}

Two of the authors (RGM, RGS) of this review paper have had the pleasure of  meeting and interacting  with the late Vadim Kuznetsov (1963--2005) at both scientif\/ic and personal levels at two recent ``Symmetry and Perturbation Theory'' Conferences held in Cala Gonone, Sardinia in
the years 2002 and 2004. Professor Kuznetsov throughout his illustrious but short career has made a major impact on the development of
the Hamilton--Jacobi theory of separation of variables as it is known to the scientif\/ic community today.

 Recall that the theory originated in the 19th century based on a method of integration of f\/inite-dimensional Hamiltonian systems. In brief, it is a procedure of f\/inding
a canonical coordinate transformation from given (position-momenta) coordinates to separable coordinates with respect to which the Hamilton--Jacobi equation associated with
a given Hamiltonian system can be integrated. In this context ``integration'' means f\/inding a complete integral satisfying a certain
non-degeneracy condition. The complete integral is usually sought under an additive separation ansatz. The principal special cases of the canonical transformation  to separable coordinates are the point transformation and the generic (non-point) transformation. The existence of separable coordinates is usually guaranteed by the existence of an additional geometric or analytic structure used to describe the dynamics of the Hamiltonian system in question. An example of separation of variables based on  the generic (non-point) canonical transformation to separable coordinates is when the Hamiltonian system under investigation is shown to
have a Lax representation. Then it may be  possible to demonstrate  that certain variables belonging to the spectral curves of the
corresponding Lax matrix serve also as the variables of separation of the Hamilton--Jacobi equation. For instance, in a recent paper \cite{vk2002} the late Professor Kuznetsov has shown that one can treat in this way both  the Kowalevski and Goryachev--Chaplygin gyrostats.

The content of this paper is two-fold. Firstly, we illustrate to the reader how to completely classify
valence two Killing tensors in the Euclidean plane relative to the action of $\mathrm{SE}(2)$. The classif\/ication is
simplif\/ied by partitioning the six-dimensional vector space of Killing tensors into submanifolds consisting of orbits
with the same dimension. Each of these submanifolds constitutes a regular foliation whose leaves are the
corresponding regular $\mathrm{SE}(2)$-orbits. By choosing appropriate transversal sections to these
leaves, the moving frame map as described by Fels and Olver \cite{Fels and Olver 1998,Fels and Olver 1999},
is used to construct suf\/f\/iciently many charts to
cover the orbits and describe the plaques. The transition maps between overlapping charts can then be used to give the
relations between the plaques. This yields an algebraic description of the leaves and hence a complete classif\/ication of
the Killing tensors. Following this, we discuss the corresponding problem for Killing two-tensors def\/ined in Euclidean
space and resulting computational dif\/f\/iculties in applying the same methodology. Secondly, given an orthogonally separable
Hamiltonian system def\/ined in the Euclidean plane by a natural Hamiltonian function of the form
\begin{gather}
H = \frac{1}{2} \big(p_1^2 + p_2^2\big) + V({\mathbf x}), \qquad {\mathbf x} = \big(x^1, x^2\big) \label{intro_ham}
\end{gather}
and the associated Killing tensor satisfying the compatibility condition with the potential, it will be shown how to use
the aforementioned classif\/ication results to easily determine:
\begin{enumerate}\itemsep=0pt
\item In which orthogonal coordinate system does the Hamiltonian separate.
\item The corresponding transformation to this orthogonally separable coordinate system.
\end{enumerate}
More specif\/ically, the outline of this paper is as follows. In Section~\ref{class} we discuss in detail the problem of
classifying Killing two-tensors by making use of the moving frame map and elementary theory associated
with regular foliations. Section~\ref{HJTs} presents a brief review of the
Hamilton--Jacobi theory of orthogonal separation of variables and its connection with the study of Killing two-tensors. The
penultimate Section~\ref{fusion} is devoted to showing how natural the problem of classifying Killing two-tensors f\/its with
the Hamilton--Jacobi theory of orthogonal separation of variables. In Section \ref{conclusion} we make f\/inal remarks.

\section[Classification]{Classif\/ication}
\label{class}

This section is a condensed version of the thesis \cite{MacThesis} and consists of two main parts. The f\/irst describes
the procedure that will be used to resolve the following two problems:
\begin{enumerate}\itemsep=0pt
\item Determine whether two elements of the vector space $\mathcal{K}^2(\mathbb{E}^2)$ of Killing-two tensors def\/ined in
the Euclidean plane are $\mathrm{SE}(2)$ equivalent.
\item Determine the transformation that takes an arbitrary element of $\mathcal{K}^2(\mathbb{E}^2)$ to its canonical or
normal form.
\end{enumerate}
The second part will present these results in their entirety.

In the next section, we show how the solution to the two problems above, i.e. the results from the thesis \cite{MacThesis},
can be used to solve the following two problems for an orthogonally separable Hamiltonian systems def\/ined in the Euclidean
plane by the natural Hamiltonian (\ref{intro_ham}):
\begin{enumerate}\itemsep=0pt
\item In which orthogonal coordinate system does the Hamiltonian separate.
\item Determine the corresponding transformation to this orthogonally separable coordinate system.
\end{enumerate}

The procedure will be described in a general manner for regular Lie group actions and will be dealt with in two components.
The f\/irst conveys how to algebraically represent the plaques of the regular foliation consisting of the orbits as leaves,
which will have constant dimension since we only consider regular actions. This gives us a local classif\/ication. The second,
will explain how these local results can be used to def\/ine the leaves globally. Subsequently, the
results of applying these ideas to classify the $\mathrm{SE}(2)$ orbits of $\mathcal{K}^2(\mathbb{E}^2)$ will be presented.

It is important to note that the action of $\mathrm{SE}(2)$ on $\mathcal{K}^2(\mathbb{E}^2)$ is not regular, i.e.\ the
dimension of the orbits may vary from point to point. Since the procedure deals only with regular actions, the initial step
will be to partition the orbits into submanifolds consisting of orbits only with the same dimension. Such a partition
will be conducted via a rank analysis on the associated distribution, in this case the Lie algebra. The action restricted
to any of these submanifolds will then be regular and the procedure may be applied.

\subsection{Local method: normalization via the moving frame map}

Here we take advantage of the normalization procedure best described in Olver's book \cite{Olver 1999}. The crucial element
is the moving frame map, which provides an explicit algorithm for constructing invariant functions for regular Lie group
actions. The normal forms are chosen as transversal sections, specif\/ically def\/ined as a regular cross-section. The
moving frame map is an equivariant map that gives the corresponding group action taking an element in a neighborhood
of the section to it's normal form. This will give us a means to compute local invariant functions by considering the
function that takes elements near the section to the their normal form as will be described in what follows.

\begin{definition}
Let $G$ be a Lie group acting semi-regularly on an $m$-dimensional manifold $M$ with $s$-dimensional orbits. A (local)
\emph{cross-section} is an ($m-s$)-dimensional submanifold $K  \subset M$ such that $K$ intersects each orbit transversally.
If the cross-section intersects each orbit at most once, then it is \emph{regular}.
\end{definition}

\begin{remark}
A \emph{coordinate cross-section} occurs when the cross-section is a level set of $s$ local coordinates on $M$. With
an appropriate choice of local coordinates, any cross-section can be made into a coordinate cross-section
(see \cite{Olver 1999}).
\end{remark}

For any point $p$ in a smooth manifold $M$, the existence of a
regular cross-section $K$ in a~neighborhood $U$ of $p$ is assured whenever
the group $G$ acts regularly (see \cite[Chapter 8]{Olver 1999}).
If in addition $G$ acts freely on $U$, then there exists a
smooth $G$-equivariant map $\psi : U \rightarrow G$ def\/ined by the condition
\begin{gather}
\psi(p) \cdot p \in K, \label{tt1}
\end{gather}
known as the \emph{moving frame map} associated with the cross-section $K$ (see Theorem 4.4 in \cite{Fels and Olver 1999} for
proof). More precisely, the moving frame map def\/ined by (\ref{tt1}) is called a \emph{right moving frame map} because it is
\emph{right equivariant}. The \emph{left moving frame map} $\widetilde{\psi}$ satisf\/ies
\begin{gather*}
\widetilde{\psi}(p)^{-1} \cdot p \in K,
\end{gather*}
and is \emph{left equivariant}. Unless otherwise stated, all moving frame maps will be right equivariant.

\begin{remark}
If the action is locally free, then the moving frame map will only be locally $G$-equivariant, i.e., in some
neighborhood $V_{e} \subset G$ of the identity.
\end{remark}

The crux of the normalization procedure lies in the following observation. Let $p \in M$ be any point whose orbit
$\mathcal{O}_p$ intersects the regular cross-section $K$ at the unique point
$k = \psi(p) \cdot p = \xi(p) \in \mathcal{O}_p \cap K$. Now, if $\widetilde{p} \neq p$ is another point in
$\mathcal{O}_p$ there is a $g \in G$ such that $\widetilde{p} = g \cdot p$. So, since $k$ is unique we have that
\begin{gather}
\xi(\widetilde{p}) = \xi(g \cdot p) = k = \xi(p), \label{tt2}
\end{gather}
i.e.\ the components $k_1 = \xi_1(p), \ldots, k_m = \xi_m(p)$ are $G$-invariant functions, $m-s$ of which are
functionally independent (since the cross-section has dimension $m-s$). Thus, a cross-section
and associated moving frame map for a regular Lie group action determines a complete set of functionally independent
invariants. We now proceed to formalize this observation into the explicit method for computing invariants known as
normalization.

Let $G$ be an $s$-dimensional Lie group with (locally) free and regular action on the $m$-dimensional
manifold $M$. Let $x = (x^1, \ldots, x^m)$ be local coordinates for $M$ in a coordinate neighborhood
$U \subseteq M$, and $g = (g^1, \ldots, g^s)$ be local coordinates for $G$ in a neighborhood\footnote{The neighborhood $V_{e}$ is chosen so that the moving frame map will be equivariant in $V_{e}$.}
$V_{e} \subseteq G$
of the identity. Now, choose a coordinate cross-section $K = \{ x^1 = c_1, \ldots, x^s = c_s \}$ def\/ined in some
neighborhood $N \subseteq U \subseteq M$. From (\ref{tt1}) and (\ref{tt2}) the moving frame map
$\psi : N \rightarrow V_e$ associated with $K$ has the following property
\begin{gather}
\psi(x) \cdot x = \big(c_1, \ldots, c_s, I_1(x), \ldots, I_{m-s}(x)\big), \label{N}
\end{gather}
where $I_1(x), \ldots, I_{m-s}(x)$ form a complete set of invariants. Equating the f\/irst $s$ components of the group
transformations $\bar{x} = g \cdot x = w(g, x)$ to the constants given by $K$, i.e.,
\begin{gather}
\bar{x}^1 = w_1(g,x) = c_1, \quad \ldots, \quad \bar{x}^s = w_s(g,x) = c_s, \label{T}
\end{gather}
must therefore implicitly def\/ine $g = \psi(x)$. The equations given by (\ref{T}) are called the
\emph{normali\-zation equations} for the coordinate cross-section $K$ and since $K$ is a well-def\/ined cross-section,
the Implicit Function Theorem implies that the group parameters in (\ref{T}) can be locally solved for in terms
of the coordinates $x$ (see \cite[Chapter~8]{Olver 1999}).

In view of (\ref{N}) it immediately follows that
\begin{gather*}
w_{s+1}(\psi(x), x) = I_1(x), \quad \ldots, \quad w_m(\psi(x), x) = I_{m-s}(x),
\end{gather*}
which are just the last $m-s$ components of $\psi(x) \cdot x$, yield a complete set of invariants.

If we have a regular cross-section and associated moving frame map $\psi$, both
of which are well-def\/ined in some neighborhood $N_p$ of a point
$p$, then the function $\xi(x) = \psi(x) \cdot x$ maps any point $x \in N_p$ to a unique
point on the cross-section. Therefore, if $\xi(x)$ maps two
points in $N_p$ to the same point on the cross-section, they are
equivalent points. That is to say, two points in $N_p$ are
equivalent if and only if evaluation of the corresponding invariant
functions at each point give the same value. Hence the local classif\/ication.

\subsection[Global classification: from plaques to leaves]{Global classif\/ication: from plaques to leaves}

Recall that in describing the classif\/ication procedure, we are only considering regular actions. As a result, all orbits
have the same dimension. We therefore consider the regular (non-singular) foliation whose leaves are the orbits of the
regular action. The idea is to initially describe the foliation locally by using the moving frame map and associated
invariant functions as distinguished charts. Slicing these charts will then yield plaques of the foliation. Each leaf
of the foliation is a union of plaques. We can determine how the plaques f\/it together to form the leaves by considering
the neighborhood relations between overlapping charts. These ideas will now be made clear.

\begin{definition} Suppose that $(U, \varphi)$ is a coordinate system on the smooth $m$-dimensional
mani\-fold~$M$ and that $d$ is an integer, $0 \leq d \leq m$. Let $(c_1, \ldots, c_m) \in \varphi(U)$ and let
\begin{gather*}
S = \{ p \in U : \varphi_i(p) = c_i, \ i = d+1, \ldots, m \}.
\end{gather*}
The subspace $S$ of $M$ together with the coordinate system
\begin{gather*}
\{ \varphi_j \ | \ S : j = 1, \ldots, d \}
\end{gather*}
is called a \emph{slice} of the coordinate system $(U, \varphi)$ and forms a manifold which is a submanifold of~$M$.
\end{definition}

A coordinate system $(U, \varphi)$ is called ``f\/lat'' for each of its slices \cite{Palais}. The normalization
procedure then, by eliciting charts that locally def\/ine the orbits as slices will locally f\/latten the orbits. The
def\/inition of a regular foliation will help precipitate this idea.

\begin{definition} A collection of arcwise connected subsets
$\mathcal{F} = \{ L_{\beta} \ | \ \beta \in B \}$ of the manifold $M$ is called a \emph{$d$-dimensional foliation}
of $M$ if it satisf\/ies the following requirements.
\begin{enumerate}\itemsep=0pt
\item Whenever $\beta, \gamma \in B$ and $\beta \neq \gamma$, \ $L_{\beta} \cap L_{\gamma} = \varnothing$.

\item The collection of subsets $\mathcal{F}$ cover $M$, i.e.\ $\underset{\beta \in B}{\cup} L_{\beta} = M$.

\item There exists a chart $(U, \varphi)$ about each $p \in M$ such that whenever $U \cap L_{\beta} \neq \varnothing$,
$\beta \in B$, each (arcwise) connected component of $\varphi(U \cap L_{\beta})$
is of the form
\begin{gather*}
\big\{ (x^1, x^2, \ldots, x^m) \in \varphi(U) \ | \ x^{d+1} = c_{d+1}, \ldots, x^m = c_m\big\}.
\end{gather*}
\end{enumerate}
\end{definition}
The subsets $L_{\beta}$ are called \emph{leaves} of the regular foliation and are $d$-dimensional submanifolds of~$M$,
since locally each is the slice of some coordinate system on~$M$. The coordinate systems that locally f\/latten the leaves
are called distinguished charts. If $L_{\beta}$ is a leaf and $(U, \varphi)$ a distinguished chart of the regular
foliation, then each arcwise connected component of $\varphi(U \cap L_{\beta})$ is called a~\emph{plaque}.

Let $G$ be a Lie transformation group acting regularly on the manifold $M$. Choose a regular cross-section $K$.
Now, apply the normalization procedure to obtain a moving frame map $\psi$ and
a complete set of invariant functions $\mathcal{I}$. If $U$ is the domain where the moving frame map $\psi$ is well-def\/ined,
then $( U, (\psi, \mathcal{I} ) )$ is a distinguished chart for the regular foliation whose leaves are the orbits of $G$ in
$M$. In particular, if $\mathcal{O}$ is an orbit such that $U \cap \mathcal{O} \neq \varnothing$, then each connected
component of $( \psi( U \cap \mathcal{O} ), \mathcal{I}(U \cap \mathcal{O}) )$ is of the form
\begin{gather}
\{ (x, y) \in ( \psi(U), \mathcal{I}(U) ) \ | \ y = \mathrm{const} \}, \label{tot}
\end{gather}
and is a plaque of the regular foliation.

The range $\mathfrak{R}$ of the action on the regular cross-section $K$ is the entire set of orbits through~$K$.
Slicing the distinguished chart as (\ref{tot}) suggests will only locally def\/ine these orbits as plaques of the regular
foliation. The remainder of the orbits through $K$ will be found in $\mathfrak{R} \setminus U$. Choosing another regular
cross-section $K'$ in the space $\mathfrak{R} \setminus U$ will then determine another such distinguished chart
$( U', (\psi', \mathcal{I}') )$ by applying again the normalization procedure to $K'$. By continuing inductively,
a suf\/f\/icient number of distinguished charts may be constructed so as to cover the orbits through $K$.

For the regular foliation whose leaves are the orbits of the action on a regular transverse section, it has
been shown how to locally def\/ine the leaves by repeatedly applying the normalization procedure to obtain a set of
distinguished charts. The leaves may be described globally by considering the neighbourhood relations between overlapping
charts. Namely, let $\varphi_{\alpha} = (x_{\alpha}, y_{\alpha}) : U_{\alpha} \rightarrow \mathbb{R}^m$,
$\varphi_{\beta} = (x_{\beta}, y_{\beta}) : U_{\beta} \rightarrow \mathbb{R}^m$ be distinguished charts for the regular
foliation on the $m$-dimensional manifold $M$. The leaves in $U_{\alpha} \cap U_{\beta}$ may then be described by
either $y_{\alpha} = \mathrm{const}$ or $y_{\beta} = \mathrm{const}$. As a result, the neighbourhood relations
$x_{\beta} = x_{\beta}( x_{\alpha}, y_{\alpha} )$ and $y_{\beta} = y_{\beta}( x_{\alpha}, y_{\alpha} )$ are such that
the $y_{\beta}$ function does not depend on $x_{\alpha}$. The corresponding function
$y_{\beta} = \tau_{\beta \alpha}( y_{\alpha} )$, called a transition function, may then be used to determine how the
plaques f\/it together to form the leaves. In particular, if we are dealing with a $d$-dimensional regular foliation and
$f_{\alpha} = \pi \circ \varphi_{\alpha} : U_{\alpha} \rightarrow \mathbb{R}^{m-d}$ where $\pi$ is the projection from
$\mathbb{R}^m$ to $\mathbb{R}^{m-d}$, then
\begin{gather*}
f_{\beta} = \tau_{\beta \alpha} ( f_{\alpha} ) \ \mathrm{in} \ U_{\alpha} \cap U_{\beta}.
\end{gather*}
That is to say, if we slice $\varphi_{\alpha}$ and $\varphi_{\beta}$ to get two plaques, then
\begin{gather*}
\{ (x, y) \in U_{\alpha} \ | \ y_{\alpha} = \tau_{\beta \alpha}( \mathrm{const} ) \} \cup \{ (x, y) \in U_{\beta} \ | \ y_{\beta} = \mathrm{const} \}
\end{gather*}
will belong to the same orbit. Therefore, the transition functions can be used to determine how
the plaques from each distinguished chart f\/it together as a union to form an orbit or leaf of the regular foliation.

\begin{remark}
For the regular foliations that arise from the action of $\mathrm{SE}(2)$ on $\mathcal{K}^2(\mathbb{E}^2)$, it was determined
that all transition functions were identities. As a result, each orbit or leaf will simply be given by taking the union of
same slice from each distinguished chart.
\end{remark}

Resolving all the leaves of the foliation with transition functions and distinguished charts will then answer the
following questions:
\begin{enumerate}\itemsep=0pt
\item Given two points $p$ and $p'$ in the manifold $M$, is there a group action between them?

\item Given a point $p \in M$, what is the group action that takes $p$ to a cross-section?
\end{enumerate}
These questions will now be addressed when $M$ is the vector space of valence-two Killing tensors
$\mathcal{K}^2( \mathbb{E}^2)$ def\/ined on the Euclidean plane, and $G$ is the two-dimensional proper Euclidean group.

\subsection[The $\mathrm{SE}(2)$-equivalence of Killing two-tensors]{The $\boldsymbol{\mathrm{SE}(2)}$-equivalence of Killing two-tensors}

The general form of the Killing two-tensor def\/ined in the Euclidean plane may be represented by the symmetric
matrix
\begin{gather}
K^{ij}(x^1, x^2) = \left(%
\begin{array}{cc}
  \alpha_1 + 2 \alpha_4 x^2 + \alpha_6 (x^2)^2 & \alpha_3 - \alpha_4 x^1 - \alpha_5 x^2 - \alpha_6 x^1 x^2 \\
  \alpha_3 - \alpha_4 x^1 - \alpha_5 x^2 - \alpha_6 x^1 x^2 & \alpha_2 + 2 \alpha_5 x^1 + \alpha_6 (x^1)^2
\end{array}%
\right). \label{MatrixKT}
\end{gather}
We wish to study the $\mathrm{SE}(2)$-invariant properties of the corresponding orthogonal webs generated
by the eigenvectors of (\ref{MatrixKT}). In order to apply the method of moving frames to this problem, we f\/irst need an
appropriate action. In particular, we are interested in the induced action of $\mathrm{SE}(2)$ on the vector space
$\mathcal{K}^2(\mathbb{E}^2)$ of valence-two Killing tensors def\/ined in the Euclidean plane. This will allow us to study
the action of $\mathrm{SE}(2)$ on the associated orthogonal webs. To obtain such an action,
we begin by applying the proper Euclidean group to the ambient manifold $\mathbb{E}^2$, which maps a point
$(x^1, x^2) \in \mathbb{E}^2$ to
\begin{gather}
\label{ActionSE2R2}
\bar{x}^1 = x^1 \cos \theta - x^2 \sin \theta + a,\qquad
\bar{x}^2 = x^1 \sin \theta + x^2 \cos \theta + b,
\end{gather}
where $\theta$, $a$, $b$ serve to parameterize the group. The push forward of (\ref{ActionSE2R2}) has the following ef\/fect
on the components $K^{ij}$,
\begin{gather*}
 \bar{K}^{ij} = K^{kl} \frac{\partial \bar{x}^i}{\partial x^k} \frac{\partial \bar{x}^j}{\partial x^l},
\end{gather*}
which induces the following transformation on the parameter space $\Sigma$, of the vector space
$\mathcal{K}^2(\mathbb{E}^2)$
\begin{gather}
\bar{\alpha}_1 = \alpha_1  \cos^2 \theta + \alpha_2  \sin^2 \theta - 2   \alpha_3  \cos   \theta  \sin   \theta - 2 b \alpha_4   \cos   \theta - 2   b   \alpha_5   \sin   \theta + \alpha_6   b^2,\nonumber\\
\bar{\alpha}_2 = \alpha_1   \sin^2 \theta + \alpha_2   \cos^2 \theta + 2   \alpha_3   \cos   \theta   \sin \theta + 2 a \alpha_4 \sin   \theta - 2   a   \alpha_5   \cos   \theta + \alpha_6   a^2,\nonumber\\
\bar{\alpha}_3 = (\alpha_1 - \alpha_2)   \sin   \theta   \cos   \theta + \alpha_3   ( \cos^2 \theta - \sin^2 \theta) + ( \alpha_4   a + \alpha_5   b )   \cos   \theta\nonumber\\
\phantom{\bar{\alpha}_3 =}{}+  ( \alpha_5   a - \alpha_4   b )   \sin   \theta - \alpha_6   a   b,\nonumber\\
\bar{\alpha}_4 = \alpha_4   \cos   \theta + \alpha_5   \sin   \theta - \alpha_6   b,\qquad
\bar{\alpha}_5 = \alpha_5   \cos   \theta - \alpha_4   \sin   \theta - \alpha_6   a,\qquad
\bar{\alpha}_6 = \alpha_6.\label{IAK2T}
\end{gather}
Note that the above transformations (\ref{IAK2T}) also appear in \cite{MST 2002JMP}.

Recall that we require the action to be regular so that we may choose a cross-section and obtain a well-def\/ined
moving frame map. As a result, we must partition the parameter space $\Sigma$ into invariant submanifolds where the
action is regular, i.e., a partition based on orbit dimension. This may be determined by a careful consideration
of the rank of the distribution for which the orbits def\/ined by (\ref{IAK2T}) are integral submanifolds.
Such a distribution is precisely the Lie algebra for the Lie group with action (\ref{IAK2T}).

The
result from \cite{The} (see \cite{MacThesis} for details) is given by Table~\ref{T7}.
\begin{table}[H]\small
  \centering
    \caption{Partition of $\Sigma$ into invariant submanifolds where the action is regular.} \label{T7}
\vspace{1mm}

\begin{tabular}{|c|c|c|}
\hline
\multicolumn{3}{|c|}{\tsep{2ex}$\begin {array}{c} \Delta_1 = (\alpha_6(\alpha_1 - \alpha_2) - \alpha_4^2 + \alpha_5^2)^2 + 4(\alpha_6 \alpha_3 + \alpha_4 \alpha_5)^2, \\
  \Delta_2 = \alpha_6, \hspace{1.0cm} \Delta_3 = (\alpha_1 - \alpha_2)^2 + 4 \alpha_3^2 \end {array}$}\bsep{2ex}\\
  \hline
 \tsep{0.5ex}  Invariant classif\/ication  & Submanifold dimension & Orbit dimension \bsep{0.5ex}\\
  \hline
\tsep{0.5ex}  $\Delta_1 \neq 0$ & 6 & 3 \bsep{0.5ex}\\
  \hline
 \tsep{0.5ex}  $\Delta_1 = 0$, $\Delta_2 \neq 0$ & 4 & 2 \bsep{0.5ex}\\
  \hline
\tsep{0.5ex}   $\Delta_1 = \Delta_2 = 0$, $\Delta_3 \neq 0$ & 3 & 1 \bsep{0.5ex}\\
  \hline
 \tsep{0.5ex}  $\Delta_1 = \Delta_2 = \Delta_3 = 0$ & 1 & 0 \bsep{0.5ex} \\
  \hline
\end{tabular}
\end{table}

\begin{remark}
The functions $\Delta_1$ and $\Delta_2$ are invariant with respect to the action (\ref{IAK2T}). The func\-tion~$\Delta_3$ is invariant
with respect to the reduced action resulting from the condition $\Delta_1 = \Delta_2 = 0$.
\end{remark}

Let $E_i \subset \Sigma$ denote the invariant submanifold given by the union of all
$i$-dimensional orbits. That is, denote
\begin{gather}
E_0 = \{ (\alpha_1, \ldots, \alpha_6) \in \Sigma \ | \ \Delta_1 = \Delta_2 = \Delta_3 = 0 \},\nonumber\\
E_1 = \{ (\alpha_1, \ldots, \alpha_6) \in \Sigma \ | \ \Delta_1 = \Delta_2 = 0, \Delta_3 \neq 0 \},\nonumber\\
E_2 = \{ (\alpha_1, \ldots, \alpha_6) \in \Sigma \ | \ \Delta_1 = 0, \Delta_2 \neq 0 \},\nonumber\\
E_3 = \{ (\alpha_1, \ldots, \alpha_6) \in \Sigma \ | \ \Delta_1 \neq 0 \}.\label{Mans}
\end{gather}

In dealing with each of the invariant submanifolds, $E_i$, only the results will be given. For the computation
see \cite{MacThesis}.

\subsubsection{The 0-dimensional orbits}

The invariant submanifold $E_0$ consists of all points in $\Sigma$ f\/ixed by the induced action
of $\mathrm{SE}(2)$. The condition $\Delta_1 = \Delta_2 = \Delta_3 = 0$ implies that $E_0$ is def\/ined by the line
$\alpha_1 = \alpha_2$, $\alpha_3 = \alpha_4 = \alpha_5 = \alpha_6 = 0$ immersed in $\Sigma$. Each point on this line
therefore takes the following form
\begin{gather}
 \left ( \begin {array}{cc}
\alpha_1 & 0 \\
 0 & \alpha_1 \end {array} \right ), \qquad \alpha_1 \in \mathbb{R}, \label{Metric}
\end{gather}
identifying with the components of the associated Killing tensor in $\mathcal{K}^2(\mathbb{E}^2)$. Fixing $\alpha_1$ in
(\ref{Metric}) therefore yields a particular $0$-dimensional orbit given by a scalar multiple of the metric.

\subsubsection{The 1-dimensional orbits}

The Killing tensors associated with points in $E_1$ generate all \emph{Cartesian webs}. Therefore, a useful choice for the
canonical forms in this case are those Cartesian webs which are aligned with the coordinate axis.
Such a regular cross-section is given by
\begin{gather}
\left ( \begin {array}{cc}
\alpha_1 & 0\\
 0 & \alpha_2 \end {array} \right ), \qquad \alpha_1 < \alpha_2. \label{CF1D}
\end{gather}
The following distinguished charts, given by Table~\ref{T8} allows us to compute the leaves (orbits) of the regular
foliation.
\begin{table}[H]\small
  \centering\caption{Distinguished charts for $E_1 \subset \Sigma \simeq \mathbb{R}^6$.} \label{T8}
  \vspace{1mm}
\begin{tabular}{|c|c|c|}
  \hline
\tsep{0.5ex}Chart  &  Coordinate function  &  Coordinate neighbourhood  \bsep{1ex}\\
  \hline
  \tsep{5ex}$(U_1, \varphi)$ & $\varphi(\alpha_1, \alpha_2, \alpha_3) = \left( \begin {array}{c} \psi_1(\alpha_1, \alpha_2, \alpha_3) \\
\noalign{\medskip} \mathcal{I}_1(\alpha_1, \alpha_2, \alpha_3) \\
\noalign{\medskip} \mathcal{I}_2(\alpha_1, \alpha_2, \alpha_3) \end {array} \right )$ & $U_1 = \{ (\alpha_1, \alpha_2, \alpha_3) \in E_1 \ | \ \alpha_3 \neq 0 \}$\bsep{5ex} \\
  \hline
 \tsep{5ex} $(U_2, \widetilde{\varphi})$ & $\widetilde{\varphi}(\alpha_1, \alpha_2, \alpha_3) = \left ( \begin {array}{c} {\psi}'_1(\alpha_1, \alpha_2, \alpha_3) \\
\noalign{\medskip} \mathcal{I}_1(\alpha_1, \alpha_2, \alpha_3) \\
\noalign{\medskip} \mathcal{I}_2(\alpha_1, \alpha_2, \alpha_3) \end {array} \right )$ & $U_2 = \{ (\alpha_1, \alpha_2, \alpha_3) \in E_1 \ | \ \alpha_3 \neq \alpha_2 \}$\bsep{5ex} \\
  \hline
\end{tabular}

\end{table}

\noindent
where
\begin{gather*}
 \psi_1(\alpha_1, \alpha_2, \alpha_3) = \arctan
\left ( \frac{\alpha_1 - \alpha_2 + \sqrt{(\alpha_1 - \alpha_2)^2 +
4 \alpha_3^2}}{2 \alpha_3} \right ),\nonumber\\
\psi_1'(\alpha_1, \alpha_2, \alpha_3) = \arctan  \left ( \frac{2 \alpha_3 +
\sqrt{(\alpha_1-\alpha_2)^2 + 4 \alpha_3^2}}{\alpha_2 - \alpha_1}
\right ),\nonumber\\  \mathcal{I}_1(\alpha_1, \alpha_2, \alpha_3) = \alpha_1 + \alpha_2,\qquad
\mathcal{I}_2(\alpha_1, \alpha_2, \alpha_3) = \alpha_3^2 - \alpha_1 \alpha_2.
\end{gather*}

Since the coordinate neighbourhoods $U_1$ and $U_2$ are not invariant, the two families of slices
\begin{gather*}
S_{\beta}^1 = \{ (\alpha_1, \alpha_2, \alpha_3) \in U_1 \ | \ \mathcal{I}_1 = \beta_1, \mathcal{I}_2 = \beta_2 \},\nonumber\\
S_{\beta}^2 = \{ (\alpha_1, \alpha_2, \alpha_3) \in U_2 \ | \ \mathcal{I}_1 = \beta_1, \mathcal{I}_2 = \beta_2 \},
\end{gather*}
must be taken together to form the leaves of the foliation. The result, gives the following representation of the leaves
\begin{gather*}
L_{\beta}^1 = S_{\beta}^1 \cup S_{\beta}^2, \qquad \beta_2 > - \beta_1^2 / 4. 
\end{gather*}

Moreover, we have the corresponding right moving frame associated with the canonical forms~(\ref{CF1D})
given by Table~\ref{T9}.
\begin{table}[H]\small
  \centering\caption{Right moving frame associated with cross-section.} \label{T9}

  \vspace{1mm}

\begin{tabular}{|c|c|}
  \hline
\tsep{0.5ex} Coordinate neighbourhood   &  Right moving frame\bsep{0.5ex} \\
  \hline
\tsep{0.5ex}  $U_1$ & $\psi_1(\alpha_1, \alpha_2, \alpha_3)$ \bsep{0.5ex}\\
  \hline
\tsep{0.5ex}  $U_2$ & $\psi'_1(\alpha_1, \alpha_2, \alpha_3) - \pi / 4$\bsep{0.5ex} \\
  \hline
\end{tabular}

\end{table}
\noindent These maps give the transformation of an arbitrary Cartesian web to it's canonical form.

\begin{example}
Consider the following two points in $E_1$
\begin{gather*}
p_1 = (\alpha_1, \alpha_2, \alpha_3) = (1, -6, 2),\qquad
p_2 = (\alpha_1, \alpha_2, \alpha_3) = (-4, 9, 1),
\end{gather*}
corresponding respectively to the Killing tensors $K$ and $\widetilde{K}$ with components
\begin{gather*}
 K^{ij} = \left ( \begin {array}{cc}
1 & 2\\
 2 & -6 \end {array} \right ), \qquad \widetilde{K}^{ij} = \left ( \begin {array}{cc}
-4 & 1\\
 1 & 9 \end {array} \right ).
\end{gather*}
To determine whether $K$ and $\widetilde{K}$ are $\mathrm{SE}(2)$-equivalent, all that is required is to check which leaf
they belong to. Namely, since
\begin{gather*}
p_1 \in L^1_{(-5, 10)}, \qquad p_2 \in L^1_{(5, 37)},
\qquad \mbox{i.e.} \qquad
L^1_{(-5, 10)} \not \equiv L^1_{(5, 37)},
\end{gather*}
the Killing tensors $K$ and $\widetilde{K}$ are not $\mathrm{SE}(2)$-equivalent (the eigenvalues of $K^{ij}$ are dif\/ferent
from those of $\widetilde{K}^{ij}$). To illustrate the transformation to canonical form, note that $K$ generates the
orthogonal coordinate web in Fig.~\ref{F7}, and $\widetilde{K}$ generates the orthogonal coordinate web in Fig.~\ref{F8}.

\begin{figure}[t]\centering
  \begin{minipage}[t]{.45\textwidth}
   \begin{center}
    \includegraphics[width=6.9cm]{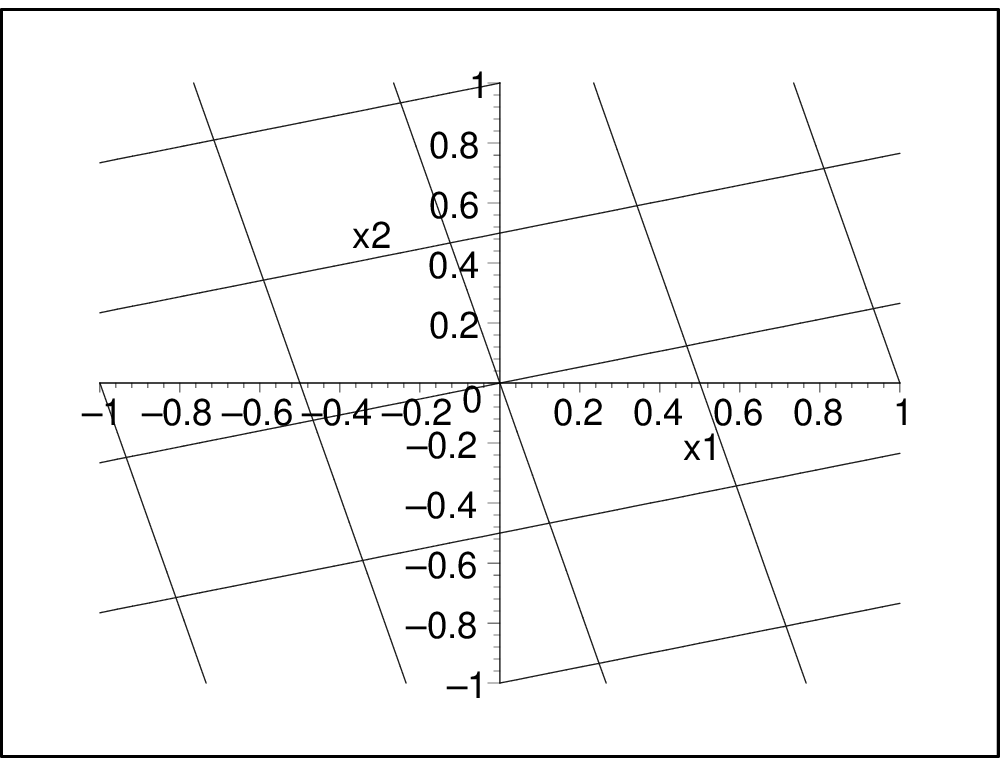}
     \caption[Cartesian web 1]{Web corresponding to $K$.}
     \label{F7}
   \end{center}
  \end{minipage}
  \qquad
  \begin{minipage}[t]{.45\textwidth}
   \begin{center}
    \includegraphics[width=6.9cm]{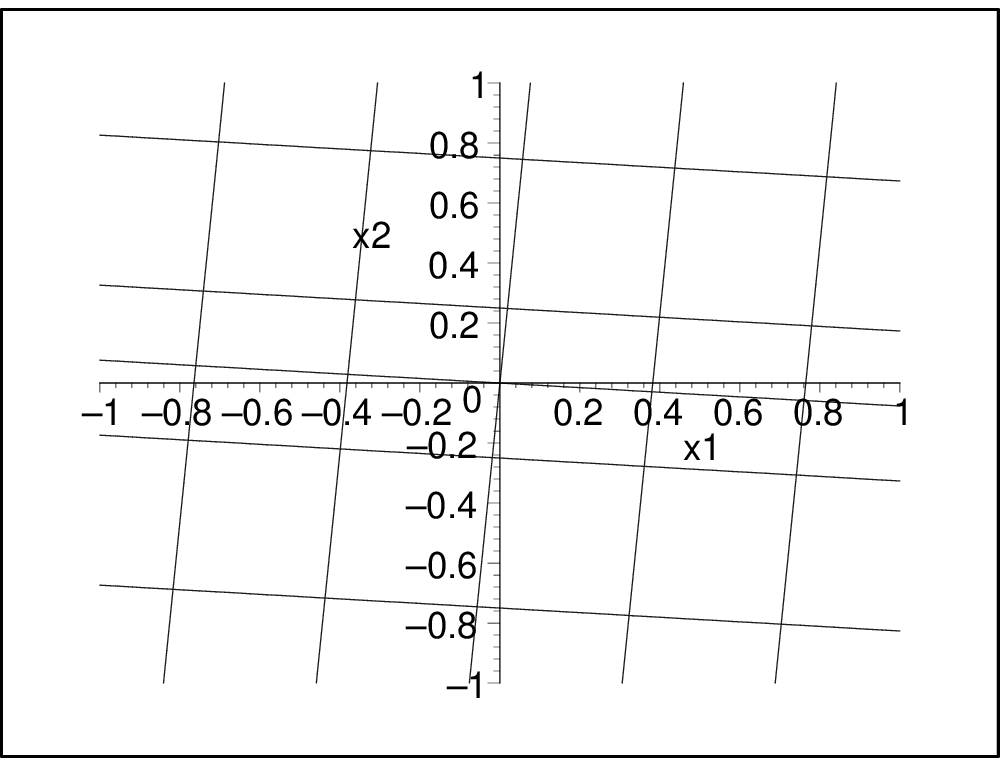}
    \caption[Cartesian web 2]{Web corresponding to $\widetilde{K}$.}
    \label{F8}
   \end{center}
  \end{minipage}
\end{figure}

\noindent Since both points $(1, -6, 2)$ and $(-4, 9, 1)$ are in $U_1$, we employ the right moving frame $\psi_1$ to obtain the
following two respective angles
\begin{gather*}
\theta_1 \approx 75 \ \mathrm{degrees},\qquad
\theta_2 \approx 4.5 \ \mathrm{degrees},
\end{gather*}
which align the orthogonal webs in Figs.~\ref{F7} and~\ref{F8} respectively with the coordinate axes for $\mathbb{E}^2$.
The components of the corresponding canonical forms in $\mathcal{K}^2(\mathbb{E}^2)$ are then
\begin{gather*}
K^{ij} \equiv \left ( \begin {array}{cc}
\displaystyle - \frac{5 + \sqrt{65}}{2} & 0\vspace{2mm}\\
 0 & \displaystyle - \frac{5 - \sqrt{65}}{2} \end {array} \right ), \qquad \widetilde{K}^{ij} \equiv \left ( \begin {array}{cc}
\displaystyle \frac{5 - \sqrt{173}}{2} & 0\vspace{2mm}\\
 0 & \displaystyle \frac{5 + \sqrt{173}}{2} \end {array} \right ),
\end{gather*}
where the entries of the canonical form for $K$ are the eigenvalues of $K^{ij}$, and the entries of the canonical form
for $\widetilde{K}$ are the eigenvalues of $\widetilde{K}^{ij}$, such that in both cases the smaller eigenvalue is in the
f\/irst row while the larger is in the second row.
\end{example}
\begin{remark}
Since the components of any other Killing tensor will not have constant eigenvalues for all $(x^1, x^2) \in \mathbb{E}^2$,
Killing tensors with parameters in $E_1$ are the only ones which generate Cartesian orthogonal webs.
\end{remark}

\subsubsection{The 2-dimensional orbits}

The Killing tensors associated with points in the invariant submanifold $E_2$ generate all the \emph{polar webs}.
In this case, we want a cross-section that corresponds to the polar webs which are aligned with the coordinate axes.
Such a regular cross-section is given by
\begin{gather*}
 \left ( \begin {array}{cc}
\alpha_1 + \alpha_6 (x^2)^2 & - \alpha_6 x^1 x^2 \vspace{1mm}\\
 - \alpha_6 x^1 x^2 & \alpha_1 + \alpha_6 (x^1)^2 \end {array} \right ), \qquad \alpha_6 \neq 0. 
\end{gather*}
The distinguished chart for $E_2$ is given below by Table~\ref{T10}.
\begin{table}[H]\small
  \centering\caption{Distinguished chart for $E_2 \subset \Sigma \simeq \mathbb{R}^6$.} \label{T10}
  \vspace{1mm}

\begin{tabular}{|c|c|c|}
  \hline
\tsep{0.5ex} Chart   &  Coordinate function  &  Coordinate neighbourhood \bsep{0.5ex} \\
  \hline
 \tsep{4ex} $(U, \varphi)$ & $\varphi(\alpha_1, \ldots, \alpha_6) = \left( \begin {array}{c} \psi(\alpha_1, \ldots, \alpha_6) \vspace{1mm}\\
 \mathcal{I}_1(\alpha_1, \ldots, \alpha_6) \vspace{1mm}\\
 \mathcal{I}_2(\alpha_1, \ldots, \alpha_6) \end {array} \right )$ & $U = E_2$ \bsep{4ex}\\
  \hline
\end{tabular}
  \end{table}

  \noindent
where
\begin{gather}
\displaystyle \psi(\alpha_1, \ldots, \alpha_6) = \left ( \frac{\alpha_5}{\alpha_6}, \ \frac{\alpha_4}{\alpha_6} \right ),\nonumber\\
\mathcal{I}_1(\alpha_1, \ldots, \alpha_6) = \alpha_6 \alpha_1 - \alpha_4^2,\qquad
\mathcal{I}_2(\alpha_1, \ldots, \alpha_6) = \alpha_6.\label{MFPo}
\end{gather}
The leaves of this foliation are then
\begin{gather}
L^2_{\beta} = \{ ( \alpha_1, \ldots, \alpha_6 ) \in E_2 \ | \ \mathcal{I}_1(\alpha_1, \ldots, \alpha_6) = \beta_1, \ \mathcal{I}_2(\alpha_1, \ldots, \alpha_6) = \beta_2 \}, \label{Levs}
\end{gather}
where $\beta_2 \neq 0$.

\begin{remark}
For any particular point $(\alpha_1, \ldots, \alpha_6) \in E_2$, the eigenvalues of the matrix represen\-ting the components
of the corresponding Killing tensor in $\mathcal{K}^2(\mathbb{E}^2)$ are given by
\begin{gather}
\label{EigsGen}
\lambda_1 = \widetilde{\alpha}_1,\qquad
 \lambda_2 = \widetilde{\alpha}_1 + \widetilde{\alpha}_6 \left ( \left ( x^1 - \frac{\alpha_5}{\alpha_6} \right )^2 + \left ( x^2 - \frac{\alpha_4}{\alpha_6} \right )^2 \right ),
\end{gather}
in the sense that the canonical form for the Killing tensor corresponding to that point
$(\alpha_1, \ldots$, $\alpha_6) \in E_2$ is given by
\begin{gather}
 \left ( \begin {array}{cc}
\widetilde{\alpha}_1 + \widetilde{\alpha}_6 (x^2)^2 & - \widetilde{\alpha}_6 x^1 x^2 \vspace{2mm}\\
 - \widetilde{\alpha}_6 x^1 x^2 & \widetilde{\alpha}_1 + \widetilde{\alpha}_6 (x^1)^2 \end {array} \right ). \label{CanPolar}
\end{gather}
\end{remark}
\begin{example}
Consider the following two points in $E_2$
\begin{gather*}
p_1 = (\alpha_1, \ldots, \alpha_6) = \left ( 2, 1, \tfrac{2}{3}, 1, 2, -3 \right ),\qquad
p_2 = (\alpha_1, \ldots, \alpha_6) = \left ( 1, -3, \tfrac{8}{3}, 2, 4, -3 \right ),
\end{gather*}
corresponding respectively to the Killing tensors $K, \widetilde{K} \in \mathcal{K}^2(\mathbb{E}^2)$ with components
\begin{gather*}
K^{ij} = \left ( \begin {array}{cc}
2 + 2 x^2 - 3 (x^2)^2 & \tfrac{2}{3} - x^1 - 2 x^2 + 3 x^1 x^2 \vspace{1mm}\\
 \tfrac{2}{3} - x^1 - 2 x^2 + 3 x^1 x^2 & 1 + 4 x^1 - 3 (x^1)^2 \end {array} \right ),\nonumber\\
\widetilde{K}^{ij} =  \left ( \begin {array}{cc}
1 + 4 x^2 - 3 (x^2)^2 & \tfrac{8}{3} - 2 x^1 - 4 x^2 + 3 x^1 x^2 \vspace{1mm}\\
\tfrac{8}{3} - 2 x^1 - 4 x^2 + 3 x^1 x^2 & -3 + 8 x^1 - 3 (x^1)^2 \end {array} \right ).
\end{gather*}
To determine whether they are $\mathrm{SE}(2)$-equivalent, check which leaf (\ref{Levs}) each belongs to. Namely, since
\begin{gather*}
p_1 \in L^2_{(-7, -3)}, \qquad p_2 \in L^2_{(-7, -3)},
\qquad \mbox{and}
\qquad
L^2_{(-7, -3)} \equiv L^2_{(-7, -3)},
\end{gather*}
the Killing tensors $K$ and $\widetilde{K}$ are $\mathrm{SE}(2)$-equivalent. The transformation to canonical form
is best illustrated with the corresponding orthogonal webs. The Killing tensor $K$ generates the orthogonal web in
Fig.~\ref{F9}, while the Killing tensor $\widetilde{K}$ generates the orthogonal web in Fig.~\ref{F10}.

\begin{figure}[t]\centering
  \begin{minipage}[t]{.45\textwidth}
   \begin{center}
    \includegraphics[width=6.9cm]{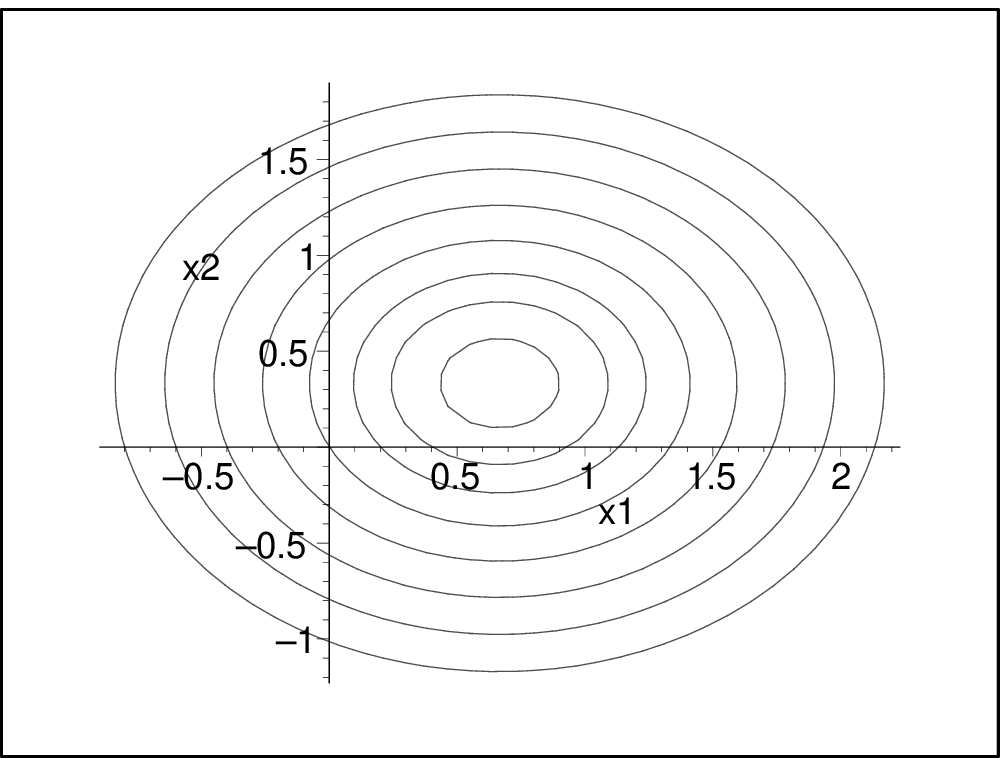}
     \caption[Polar web 1]{Web corresponding to $K$.}
     \label{F9}
   \end{center}
  \end{minipage}
  \qquad
  \begin{minipage}[t]{.45\textwidth}
   \begin{center}
    \includegraphics[width=6.9cm]{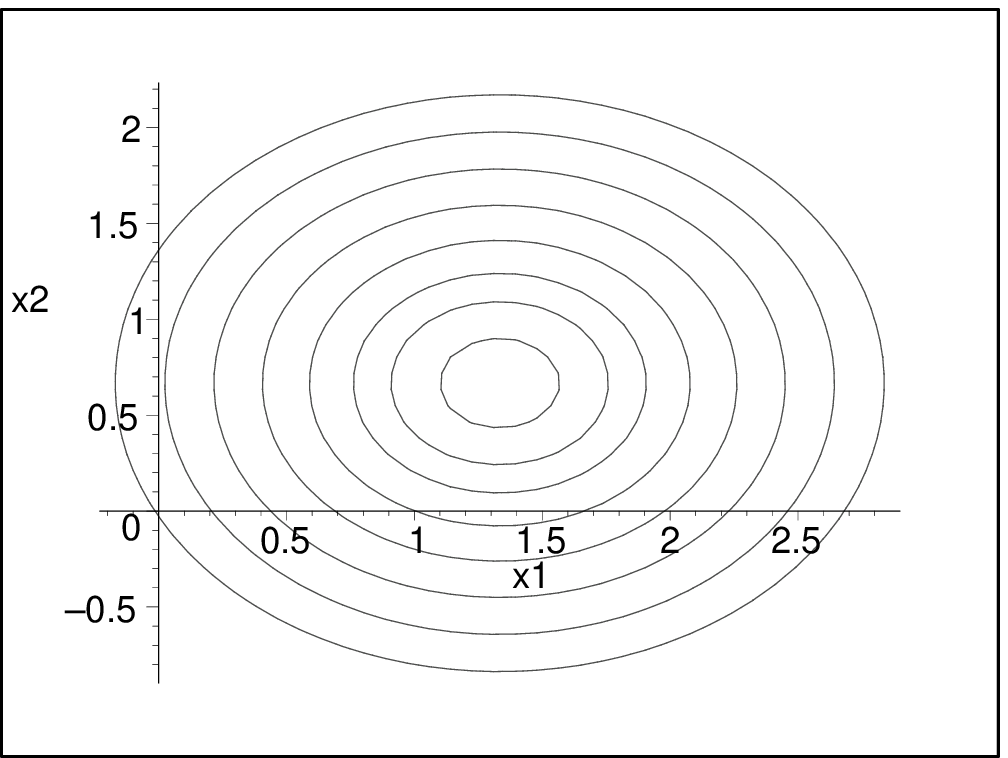}
    \caption[Polar web 2]{Web corresponding to $\widetilde{K}$.}
    \label{F10}
   \end{center}
  \end{minipage}
\end{figure}

 Employing the right moving frame (\ref{MFPo}) and substituting into the action (\ref{ActionSE2R2}) with $\theta = 0$ yields
the following transformations,
\begin{gather*}
 \bar{x}^1 = x^1 - \tfrac{2}{3},\qquad
 \bar{x}^2 = x^2 - \tfrac{1}{3},
\end{gather*}
which maps the singular point of the orthogonal web in Fig.~\ref{F9} to the origin, and
\begin{gather*}
\bar{x}^1 = x^1 - \tfrac{4}{3},\qquad
\bar{x}^2 = x^2 - \tfrac{2}{3},
\end{gather*}
which maps the singular point of the orthogonal web in Fig.~\ref{F10} to the origin. Furthermore, utilizing
(\ref{CanPolar}) and (\ref{EigsGen}) yields the components of the canonical form for $K$ and $\widetilde{K}$
\begin{gather*}
K \equiv \widetilde{K} \equiv \left ( \begin {array}{cc}
 \tfrac{7}{3} - 3 (x^2)^2 &  3 x^1 x^2 \vspace{1mm}\\
 3 x^1 x^2 &  \tfrac{7}{3} - 3 (x^1)^2 \end {array} \right ).
\end{gather*}
\end{example}

\subsubsection{The 3-dimensional orbits}

The Killing tensors that correspond to points in the invariant submanifold $E_3$ generate either
a \emph{parabolic web} or an \emph{elliptic-hyperbolic web}. We may partition $E_3$ further by taking advantage
of the fact that Killing tensors in $E_3$ with $\alpha_6 = 0$ have one singular point, while those
in $E_3$ with $\alpha_6 \neq 0$ have two. Namely, set
\begin{gather}
E_3^1 = \{ ( \alpha_1, \alpha_2, \ldots, \alpha_6 ) \in E_3 \ | \ \Delta_2 = \alpha_6 = 0 \},\nonumber\\
E_3^2 = \{ ( \alpha_1, \alpha_2, \ldots, \alpha_6 ) \in E_3 \ | \ \Delta_2 = \alpha_6 \neq 0 \},\label{EHPA}
\end{gather}
so that $E_3^1$ consists of all parabolic webs and $E_3^2$ consists of all elliptic-hyperbolic webs.

{\textbf{Parabolic webs.}}
Considering f\/irst the family of parabolic webs, we choose a regular cross-section so that the singular point is at the
origin and the web is rotationally aligned with the axes. Such a collection of canonical forms is given by,
\begin{gather}
\label{ConfPa}
\left ( \begin {array}{cc}
\alpha_1 & - \alpha_5 x^2 \vspace{1mm}\\
 - \alpha_5 x^2 & \alpha_1 + 2 \alpha_5 x^1 \end {array} \right ), \qquad \alpha_5 > 0.
\end{gather}
The distinguished charts for $E_3^1$ is given by Table~\ref{T11} below.
\begin{table}[H]\small
  \centering \caption{Distinguished charts for $E_3^1 \subset \Sigma \simeq \mathbb{R}^6$.} \label{T11}

  \vspace{1mm}

\begin{tabular}{|c|c|c|}
  \hline
\tsep{0.5ex} Chart   &  Coordinate function  &  Coordinate neighbourhood \bsep{0.5ex} \\
  \hline
\tsep{4ex}  $(U_1, \varphi)$ & $\varphi(\alpha_1, \ldots, \alpha_6) = \left( \begin {array}{c} \psi(\alpha_1, \ldots, \alpha_6) \\
 \mathcal{I}_1(\alpha_1, \ldots, \alpha_6) \\
 \mathcal{I}_2(\alpha_1, \ldots, \alpha_6) \end {array} \right )$ &
 $U_1 = \{ ( \alpha_1, \ldots, \alpha_6 ) \in E_3^1 \ | \ \alpha_5 > 0 \}$ \bsep{4ex}\\
  \hline
\tsep{4ex}   $(U_2, \varphi)$ & $\varphi(\alpha_1, \ldots, \alpha_6) = \left( \begin {array}{c} \psi(\alpha_1, \ldots, \alpha_6) \\
 \mathcal{I}_1(\alpha_1, \ldots, \alpha_6)\\
\mathcal{I}_2(\alpha_1, \ldots, \alpha_6) \end {array} \right )$ & $U_2 = \{ ( \alpha_1, \ldots, \alpha_6 ) \in E_3^1 \ | \ \alpha_5 < 0 \}$\bsep{4ex} \\
  \hline
\end{tabular}
 \end{table}

\noindent
where
\begin{gather}
\psi(\alpha_1, \ldots, \alpha_6) = \left(%
\begin{array}{c}
  \theta \\
  a \\
  b
\end{array}
\right) = \left(
\begin{array}{c}
  - \arctan  ( \alpha_4 / \alpha_5 ) \vspace{2mm}\\
  \displaystyle \frac{\alpha_5 \sqrt{\frac{\alpha_4^2 + \alpha_5^2}{\alpha_5^2}} (( \alpha_1 - \alpha_2 )( \alpha_4^2 - \alpha_5^2 ) - 4 \alpha_3 \alpha_4 \alpha_5 )  }{2 (\alpha_4^2 + \alpha_5^2)^{2}} \vspace{2mm}\\
  \displaystyle \frac{\alpha_5 \sqrt{\frac{\alpha_4^2 + \alpha_5^2}{\alpha_5^2}} (\alpha_1 \alpha_4 \alpha_5 - \alpha_2 \alpha_4 \alpha_5 - \alpha_3 \alpha_5^2 + \alpha_3 \alpha_4^2) }{(\alpha_4^2 + \alpha_5^2)^{2}}
\end{array}
\right),\nonumber\\
\mathcal{I}_1(\alpha_1, \ldots, \alpha_6) = \alpha_4^2 + \alpha_5^2,\qquad
\mathcal{I}_2(\alpha_1, \ldots, \alpha_6) = 2 \alpha_3 \alpha_4 \alpha_5 + \alpha_1 \alpha_5^2 + \alpha_2 \alpha_4^2.
\label{MF1Pa}
\end{gather}

In this case, the two families of slices
\begin{gather}
\widetilde{S}^1_{\beta} =\{ (\alpha_1, \ldots, \alpha_6) \in U_1 \ | \ \mathcal{I}_1(\alpha_1, \ldots, \alpha_6) = \beta_1, \mathcal{I}_2(\alpha_1, \ldots, \alpha_6) = \beta_2 \},\nonumber\\
\widetilde{S}^2_{\beta} =\{ (\alpha_1, \ldots, \alpha_6) \in U_2 \ | \ \mathcal{I}_1(\alpha_1, \ldots, \alpha_6) = \beta_1, \mathcal{I}_2(\alpha_1, \ldots, \alpha_6) = \beta_2 \},\label{SlicesPa}
\end{gather}
must be taken together to form the leaves $\widetilde{L}^3_{\beta}$ of the regular foliation. Namely,
\begin{gather}
\widetilde{L}^3_{\beta} = \widetilde{S}^1_{\beta} \cup \widetilde{S}^2_{\beta} \label{LevsPa}
\end{gather}
for all $(\beta_1, \beta_2) \in \mathcal{I}_1( E_3^1 ) \times \mathcal{I}_2( E_3^1 )$.

Consult Table~\ref{T12} for the moving frame that maps an arbitrary
element in $E_3^1$ to the cross-section (\ref{ConfPa}). See (\ref{MF1Pa}) for $(\theta, a, b)$.
\begin{table}[H]\small
  \centering \caption{Right moving frame for the global regular cross-section (\ref{ConfPa}).} \label{T12}

  \vspace{1mm}

\begin{tabular}{|c|c|}
  \hline
\tsep{0.5ex} \raisebox{-1.5ex}[0pt][0pt]{Coordinate neighbourhood}   &  Right moving frame:  \\
   & See (\ref{MF1Pa}) for $(\theta, a, b)$ \bsep{0.5ex} \\
  \hline
\tsep{3ex}  $(\alpha_1, \ldots, \alpha_6) \in U_1$ & $ \left(\!\! \begin {array}{c} \theta \\
 a \\
 b  \end {array}\!\! \right )$ \bsep{3ex}\\
  \hline
\tsep{3ex}  $(\alpha_1, \ldots, \alpha_6) \in U_2$ & $\left( \!\!\begin {array}{c} \theta \pm \pi \\
 - a \\
 - b \end {array}\!\! \right )$ \bsep{3ex}\\
  \hline
\end{tabular}

\end{table}
In addition, the canonical form for an arbitrary point in $E_3^1$ may be determined by the invariant functions (\ref{MF1Pa}).
That is, the components of the canonical form for a Killing tensor corresponding to an arbitrary point in $E_3^1$
is given by
\begin{gather}
\label{CanPa}
\left ( \begin {array}{cc}
\displaystyle \frac{\mathcal{I}_2}{\mathcal{I}_1} & - \sqrt{ \mathcal{I}_1 }   x^2 \vspace{1mm}\\
 - \sqrt{ \mathcal{I}_1 }   x^2 & \displaystyle \frac{\mathcal{I}_2}{\mathcal{I}_1} + 2 \sqrt{ \mathcal{I}_1 }   x^1 \end {array} \right ).
\end{gather}

\begin{example}
Consider the following two points in $E_3^1$
\begin{gather*}
p_1 = (\alpha_1, \ldots, \alpha_6) = (1, -3, 5, 1, 2, 0),\qquad
p_2 = (\alpha_1, \ldots, \alpha_6) = (-2, 5, 7, 0, -1, 0),
\end{gather*}
corresponding respectively to the Killing tensors $K$ and $\widetilde{K}$ with components
\begin{gather*}
K^{ij} = \left ( \begin {array}{cc}
2 + 2 x^2 - 3 (x^2)^2 & \tfrac{2}{3} - x^1 - 2 x^2 + 3 x^1 x^2 \vspace{1mm}\\
\tfrac{2}{3} - x^1 - 2 x^2 + 3 x^1 x^2 & 1 + 4 x^1 - 3 (x^1)^2 \end {array} \right ),\nonumber\\
\widetilde{K}^{ij} =  \left ( \begin {array}{cc}
1 + 4 x^2 - 3 (x^2)^2 &  \tfrac{8}{3} - 2 x^1 - 4 x^2 + 3 x^1 x^2 \vspace{1mm}\\
 \tfrac{8}{3} - 2 x^1 - 4 x^2 + 3 x^1 x^2 & -3 + 8 x^1 - 3 (x^1)^2 \end {array} \right ).
\end{gather*}
To determine whether $K$ and $\widetilde{K}$ belong to the same equivalence class, utilize (\ref{SlicesPa}) and
(\ref{LevsPa}) to f\/ind out which leaf each belongs to. Namely, since
\begin{gather*}
p_1 \in \widetilde{L}^3_{(5, 21)}, \hspace{1.0cm} p_2 \in \widetilde{L}^3_{(1, -2)},
\qquad \mbox{and}\qquad
\widetilde{L}^3_{(5,   21)} \not \equiv \widetilde{L}^3_{(1,   -2)},
\end{gather*}
the Killing tensors $K$ and $\widetilde{K}$ are not $\mathrm{SE}(2)$-equivalent. To illustrate the transformation to
canonical form, consider the web generated by $K$ (see Fig.~\ref{F11}) and the web generated by $\widetilde{K}$
(see Fig.~\ref{F12}).

\begin{figure}[t]\centering
  \begin{minipage}[t]{.45\textwidth}
   \begin{center}
    \includegraphics[width=6.9cm]{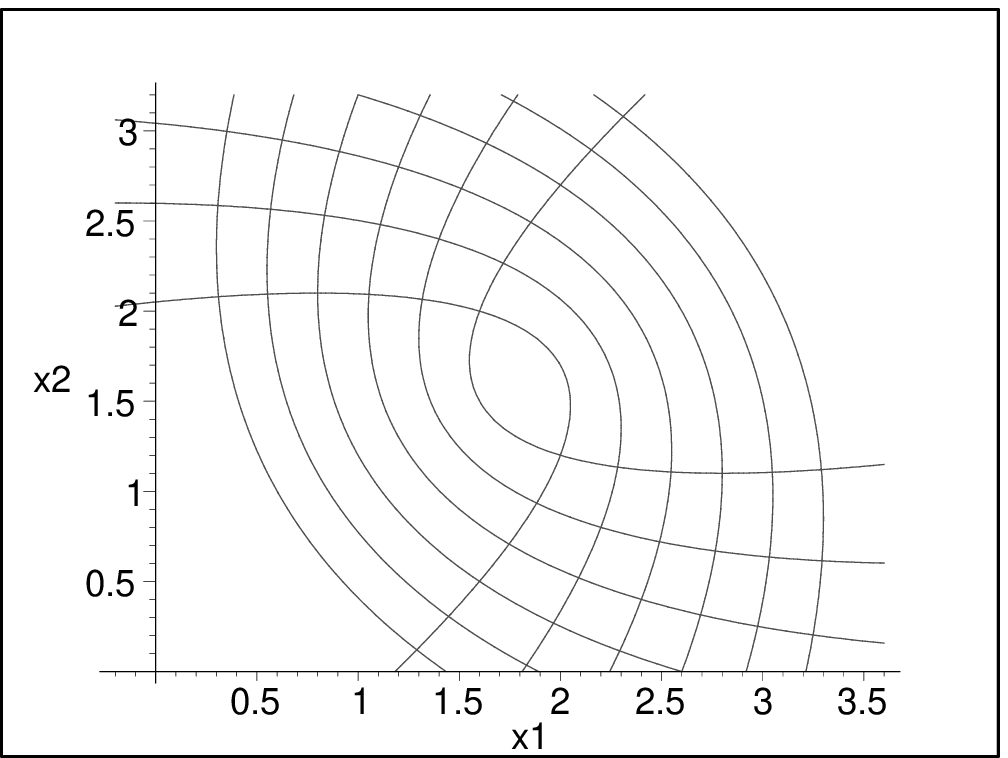}
     \caption[Parabolic web 1]{Web corresponding to $K$.}
     \label{F11}
   \end{center}
  \end{minipage}
  \qquad
  \begin{minipage}[t]{.45\textwidth}
   \begin{center}
    \includegraphics[width=6.9cm]{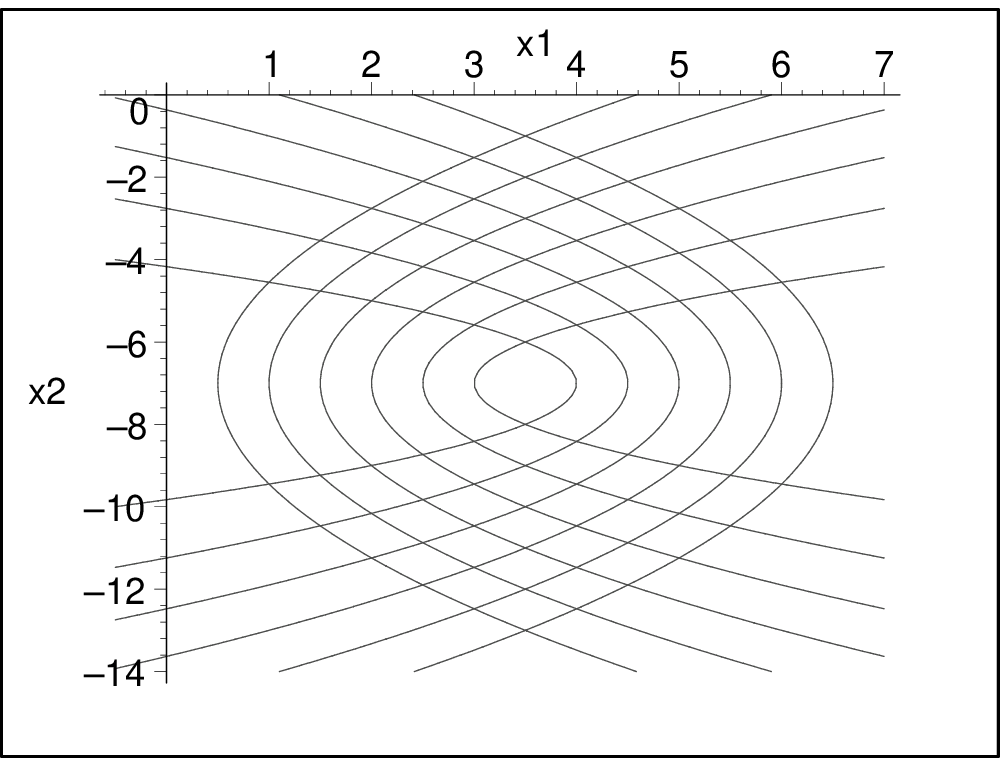}
    \caption[Parabolic web 2]{Web corresponding to $\widetilde{K}$.}
    \label{F12}
   \end{center}
  \end{minipage}
\end{figure}

The right moving frame from Table~\ref{T12} immediately tells us that the map which takes the web in Fig.~\ref{F11} to
its canonical form requires the following rotation and translation
\begin{gather}
\theta = - \arctan \left(\tfrac 12\right) \ \mathrm{radians} \approx -27 \
\mathrm{degrees},\qquad   a = - \tfrac{26
\sqrt{5}}{25},\qquad  b = - \tfrac{7 \sqrt{5}}{25},\label{MFEx1Pa}
\end{gather}
where $a$ is a horizontal translation and $b$ is a vertical translation (applied after the rotation). Substituting
(\ref{MFEx1Pa}) into the action (\ref{ActionSE2R2}) yields the following transformation
\begin{gather*}
\bar{x}^1 = \tfrac{2}{\sqrt{5}} x^1 + \tfrac{1}{\sqrt{5}} x^2 - \tfrac{26 \sqrt{5}}{25},\qquad
\bar{x}^2 = - \tfrac{1}{\sqrt{5}} x^1 + \tfrac{2}{\sqrt{5}} x^2 - \tfrac{7 \sqrt{5}}{25},
\end{gather*}
which maps the singular point to the origin and aligns the web from Fig.~\ref{F11} with the coordinate axes for the
ambient manifold $\mathbb{E}^2$. Similarly, applying the right moving frame to the web in Fig.~\ref{F12} yields the
following rotation and translation
\begin{gather}
\label{MFEx2Pa}
\theta = \pm \ \pi \ \mathrm{radians},\qquad
 a = \tfrac{7}{2},\qquad
 b = - 7.
\end{gather}
Substituting (\ref{MFEx2Pa}) into the action (\ref{ActionSE2R2}) then gives the following transformation
\begin{gather*}
\bar{x}^1 = - x^1 + \tfrac{7}{2},\qquad
 \bar{x}^2 = - x^2 - 7,
\end{gather*}
which maps the singular point to the origin and aligns the web from Fig.~\ref{F12} with the coordinate axes for the
ambient manifold $\mathbb{E}^2$.

Implementing formula (\ref{CanPa}) gives the following canonical forms on the cross-section (\ref{ConfPa}) for the components
of the associated Killing tensors $K$ and $\widetilde{K}$
\begin{gather*}
\label{CanExPa}
\begin{array}{l}
K^{ij} \equiv \left ( \begin {array}{cc}
\displaystyle \tfrac{21}{5} & - \sqrt{5} x^2 \vspace{1mm}\\
 - \sqrt{5} x^2 & \displaystyle \tfrac{21}{5} + 2 \sqrt{5} x^1 \end {array} \right ), \qquad
  \widetilde{K}^{ij} \equiv  \left ( \begin {array}{cc}
-2 & - x^2 \\
\noalign{\medskip} - x^2 & -2 + 2 x^1 \end {array} \right ).
\end{array}
\end{gather*}
\end{example}

{\textbf{Elliptic-hyperbolic webs.}}
For the family of elliptic-hyperbolic webs, the canonical forms should be those webs aligned with the coordinate axes
for the ambient manifold $\mathbb{E}^2$. That is, we choose those webs whose singular points are on the horizontal axis
and equi-distant from the origin. In this regard, the regular cross-section
\begin{gather}
\label{CanEH}
\left ( \begin {array}{cc}
\alpha_1 + \alpha_6 (x^2)^2 & - \alpha_6 x^1 x^2 \\
 - \alpha_6 x^1 x^2 & \alpha_2 + \alpha_6 (x^1)^2 \end {array} \right ), \qquad \alpha_6 ( \alpha_1 - \alpha_2 ) > 0,
\end{gather}
satisf\/ies the desired criteria.

The distinguished charts for the invariant submanifold $E_3^2$ is given by Table~\ref{T13} below.
\begin{table}[H]\small
  \centering   \caption{Distinguished charts for $E_3^2 \subset \Sigma \simeq \mathbb{R}^6$.} \label{T13}

  \vspace{1mm}

  \begin{tabular}{|c|c|c|}
      \hline
      \multicolumn{3}{|c|}{\tsep{0.5ex}$\iota_1 = \alpha_6 (\alpha_1 - \alpha_2) - \alpha_4^2 + \alpha_5^2$, \qquad $\iota_2 = \alpha_3   \alpha_6 + \alpha_4   \alpha_5$\bsep{0.5ex}} \\
      \hline
\tsep{0.5ex}  Chart   &  Coordinate function  &  Coordinate Neighbourhood  \bsep{0.5ex}\\
  \hline
\tsep{5ex}  $(U_1, \varphi)$ & $\varphi(\alpha_1, \ldots, \alpha_6) = \left( \begin {array}{c} \psi_1(\alpha_1, \ldots, \alpha_6) \\
 \mathcal{I}_1(\alpha_1, \ldots, \alpha_6) \\
 \mathcal{I}_2(\alpha_1, \ldots, \alpha_6) \\
 \mathcal{I}_3(\alpha_1, \ldots, \alpha_6) \end {array} \right )$ & $U_1 = \{ ( \alpha_1, \ldots, \alpha_6 ) \in E_3^2 \ | \ \iota_1 > 0 \}$ \bsep{5ex}\\
  \hline
 \tsep{5ex} $(U_2, \varphi)$ & $\varphi(\alpha_1, \ldots, \alpha_6) = \left( \begin {array}{c} \psi_1(\alpha_1, \ldots, \alpha_6) \\
 \mathcal{I}_1(\alpha_1, \ldots, \alpha_6) \\
 \mathcal{I}_2(\alpha_1, \ldots, \alpha_6) \\
\mathcal{I}_3(\alpha_1, \ldots, \alpha_6) \end {array} \right )$ & $U_2 = \{ ( \alpha_1, \ldots, \alpha_6 ) \in E_3^2 \ | \ \iota_1 < 0 \}$\bsep{5ex} \\
  \hline
\tsep{5ex}  $(U_3, \widetilde{\varphi})$ & $\varphi(\alpha_1, \ldots, \alpha_6) = \left( \begin {array}{c} \psi_2(\alpha_1, \ldots, \alpha_6) \\
 \mathcal{I}_1(\alpha_1, \ldots, \alpha_6) \\
 \mathcal{I}_2(\alpha_1, \ldots, \alpha_6) \\
 \mathcal{I}_3(\alpha_1, \ldots, \alpha_6) \end {array} \right )$ & $U_3 = \{ ( \alpha_1, \ldots, \alpha_6 ) \in E_3^2 \ | \ \iota_2 > 0 \}$ \bsep{5ex}\\
  \hline
\tsep{5ex}  $(U_4, \widetilde{\varphi})$ & $\varphi(\alpha_1, \ldots, \alpha_6) = \left( \begin {array}{c} \psi_2(\alpha_1, \ldots, \alpha_6) \\
 \mathcal{I}_1(\alpha_1, \ldots, \alpha_6) \\
 \mathcal{I}_2(\alpha_1, \ldots, \alpha_6) \\
 \mathcal{I}_3(\alpha_1, \ldots, \alpha_6) \end {array} \right )$ & $U_4 = \{ ( \alpha_1, \ldots, \alpha_6 ) \in E_3^2 \ | \ \iota_2 < 0 \}$ \bsep{5ex} \\
  \hline
    \end{tabular}
\end{table}

\noindent
where
\begin{gather}
\psi_1(\alpha_1, \ldots, \alpha_6) = \left(%
\begin{array}{c}
  \theta_1 \\
  a_1 \\
  b_1 \\
\end{array}%
\right) = \left(%
\begin{array}{c}
  \displaystyle - \frac{1}{2}  \arctan  \left ( \frac{2( \alpha_3 \alpha_6 + \alpha_4 \alpha_5 )}{ \alpha_6(\alpha_1 - \alpha_2) - \alpha_4^2 + \alpha_5^2 } \right ) \vspace{1mm}\\
  \displaystyle \frac{\alpha_5   \cos   \theta_1 - \alpha_4   \sin   \theta_1}{\alpha_6} \vspace{1mm}\\
  \displaystyle \frac{\alpha_4   \cos   \theta_1 + \alpha_5   \sin   \theta_1}{\alpha_6}
\end{array}%
\right),\nonumber\\
\label{MF2EH}
\psi_2(\alpha_1, \ldots, \alpha_6) = \left(%
\begin{array}{c}
  \theta_2 \\
  a_2 \\
  b_2
\end{array}%
\right) = \left(%
\begin{array}{c}
  \displaystyle \frac{1}{2}  \arctan  \left ( \frac{ \alpha_6(\alpha_1 - \alpha_2) - \alpha_4^2 + \alpha_5^2 }{2( \alpha_3 \alpha_6 + \alpha_4 \alpha_5 )} \right ) \vspace{1mm}\\
  \displaystyle \frac{\alpha_5   \cos   \theta_2 - \alpha_4   \sin   \theta_2}{\alpha_6} \vspace{1mm}\\
  \displaystyle \frac{\alpha_4   \cos   \theta_2 + \alpha_5   \sin   \theta_2}{\alpha_6}
\end{array}%
\right),\\
\mathcal{I}_1(\alpha_1, \ldots, \alpha_6) = \alpha_6,\qquad
\mathcal{I}_2(\alpha_1, \ldots, \alpha_6) = \alpha_6 ( \alpha_1 + \alpha_2 ) - \alpha_4^2 - \alpha_5^2,\nonumber\\
\mathcal{I}_3(\alpha_1, \ldots, \alpha_6) = \alpha_6 ( \alpha_3^2 - \alpha_1 \alpha_2 ) + \alpha_4^2 \alpha_2 + 2 \alpha_3 \alpha_4 \alpha_5 + \alpha_1 \alpha_5^2.\nonumber
\end{gather}

Each leaf of the regular foliation is a union of the slices
\begin{gather}
S^1_{\beta} = \{ (\alpha_1, \ldots, \alpha_6) \in U_1 \ | \ \mathcal{I}_1 = \beta_1, \mathcal{I}_2 = \beta_2, \mathcal{I}_3 = \beta_3 \},\nonumber\\
S^2_{\beta} = \{ (\alpha_1, \ldots, \alpha_6) \in U_2 \ | \ \mathcal{I}_1 = \beta_1, \mathcal{I}_2 = \beta_2, \mathcal{I}_3 = \beta_3 \},\nonumber\\
S^3_{\beta} = \{ (\alpha_1, \ldots, \alpha_6) \in U_3 \ | \ \mathcal{I}_1 = \beta_1, \mathcal{I}_2 = \beta_2, \mathcal{I}_3 = \beta_3 \},\nonumber\\
S^4_{\beta} = \{ (\alpha_1, \ldots, \alpha_6) \in U_4 \ | \ \mathcal{I}_1 = \beta_1, \mathcal{I}_2 = \beta_2, \mathcal{I}_3 = \beta_3 \},\label{SlicesEH1}
\end{gather}
In particular, the leaves, indexed by all $\beta = (\beta_1, \beta_2, \beta_3)$ in
$\mathcal{I}_1(E_3^2) \times \mathcal{I}_2(E_3^2) \times \mathcal{I}_3(E_3^2)$, are given~by
\begin{gather}
L^3_{\beta} = \bigcup_{i=1}^4 S^i_{\beta}. \label{LevsEH}
\end{gather}

For the right moving frame giving the transformation that takes a Killing tensor to its cano\-nical form (\ref{CanEH}),
consult Table~\ref{T14} below.
\begin{table}[H]\small
  \centering \caption{Right moving frame for the global regular cross-section (\ref{CanEH}).} \label{T14}
  \vspace{1mm}

\begin{tabular}{|c|c|}
  \hline
 \tsep{0.5ex}  &  Right moving frame:  \\
 \raisebox{1.5ex}[0pt][0pt]{Coordinate neighbourhood}  & see (\ref{MF2EH}) for $(\theta_i, a_i, b_i)$, $i = 1, 2$ \bsep{0.5ex}\\
  \hline
 \tsep{4ex} $(\alpha_1, \ldots, \alpha_6) \in U_1$ & $\displaystyle \left( \begin {array}{c} \theta \\
 a \\
 b  \end {array} \right ) = \left( \begin {array}{c} \theta_1 \\
 a_1 \\
 b_1  \end {array} \right )$  \bsep{4ex} \\
  \hline
 \tsep{4ex}  $(\alpha_1, \ldots, \alpha_6) \in U_2$ & $\displaystyle \left( \begin {array}{c} \theta \\
 a \\
 b  \end {array} \right ) = \left( \begin {array}{c} \theta_1 \pm \pi / 2 \\
 \mp b_1 \\
 \pm a_1 \end {array} \right )$ \bsep{4ex}\\
  \hline
 \tsep{4ex}  $(\alpha_1, \ldots, \alpha_6) \in U_3$ & $\displaystyle \left( \begin {array}{c} \theta \\
 a \\
 b  \end {array} \right ) = \left( \begin {array}{c} \theta_2 - \pi / 4 \\
 \displaystyle \frac{a_2}{\sqrt{2}} + \frac{b_2}{\sqrt{2}} \\
 \displaystyle - \frac{a_2}{\sqrt{2}} + \frac{b_2}{\sqrt{2}} \end {array} \right )$ \bsep{4ex}\\
  \hline
 \tsep{4ex}  $(\alpha_1, \ldots, \alpha_6) \in U_4$ & $\displaystyle \left( \begin {array}{c} \theta \\
 a \\
 b  \end {array} \right ) = \left( \begin {array}{c} \theta_2 + \pi / 4 \\
 \displaystyle \frac{a_2}{\sqrt{2}} - \frac{b_2}{\sqrt{2}} \\
 \displaystyle \frac{a_2}{\sqrt{2}} + \frac{b_2}{\sqrt{2}} \end {array} \right )$  \bsep{4ex}\\
  \hline
\end{tabular}
 \end{table}

Moreover, the formula for the canonical form of an arbitrary point in $E_3^2$ is
\begin{gather}
P^+ = \left ( \frac{\mathcal{I}_2}{2 \mathcal{I}_1 } + \sqrt{ \frac{\mathcal{I}_3}{\mathcal{I}_1} + \left ( \frac{\mathcal{I}_2}{2 \mathcal{I}_1} \right )^2 }, \ \frac{\mathcal{I}_2}{2 \mathcal{I}_1 } - \sqrt{ \frac{\mathcal{I}_3}{\mathcal{I}_1} + \left ( \frac{\mathcal{I}_2}{2 \mathcal{I}_1} \right )^2 }, 0, 0, 0, \mathcal{I}_1  \right ), \label{PPlus}
\end{gather}
whenever $\mathcal{I}_1 > 0$, and the point
\begin{gather}
P^- = \left ( \frac{\mathcal{I}_2}{2 \mathcal{I}_1 } - \sqrt{ \frac{\mathcal{I}_3}{\mathcal{I}_1} + \left ( \frac{\mathcal{I}_2}{2 \mathcal{I}_1} \right )^2 }, \ \frac{\mathcal{I}_2}{2 \mathcal{I}_1 } + \sqrt{ \frac{\mathcal{I}_3}{\mathcal{I}_1} + \left ( \frac{\mathcal{I}_2}{2 \mathcal{I}_1} \right )^2 }, 0, 0, 0, \mathcal{I}_1  \right ), \label{PMinus}
\end{gather}
whenever $\mathcal{I}_1 < 0$.
\begin{example}
Consider the following two points in $E_3^2$
\begin{gather*}
p_1 = (\alpha_1, \ldots, \alpha_6) = (2, 1, 0, 1, 1, 4),\qquad
p_2 = (\alpha_1, \ldots, \alpha_6) = (2, 1, 0, 1, 1, -4),
\end{gather*}
corresponding respectively to the Killing tensors $K$ and $\widetilde{K}$ with components
\begin{gather*}
K^{ij} = \left ( \begin {array}{cc}
2 + 2 x^2 + 4 (x^2)^2 & - x^1 - x^2 - 4 x^1 x^2 \vspace{1mm}\\
 - x^1 - x^2 - 4 x^1 x^2 & 1 + 2 x^1 + 4 (x^1)^2 \end {array} \right ),\nonumber\\
\widetilde{K}^{ij} = \left ( \begin {array}{cc}
2 + 2 x^2 - 4 (x^2)^2 & - x^1 - x^2 + 4 x^1 x^2 \vspace{1mm}\\
 - x^1 - x^2 + 4 x^1 x^2 & 1 + 2 x^1 - 4 (x^1)^2 \end {array} \right ).
\end{gather*}
To determine whether $K$ and $\widetilde{K}$ are $\mathrm{SE}(2)$-equivalent, use the formulae (\ref{SlicesEH1}) for the
slices to ascertain which leaf, using (\ref{LevsEH}), each belongs to. That is to say, since
\begin{gather*}
p_1 \in L^3_{(4, 10, -5)}, \qquad p_2 \in L^3_{(-4, -14, 11)},
\qquad \mbox{and}\qquad
L^3_{(4, 10, -5)} \not \equiv L^3_{(-4, -14, 11)},
\end{gather*}
the Killing tensors $K$ and $\widetilde{K}$ are not $\mathrm{SE}(2)$-equivalent. To illustrate the transformation to
canonical form, it is useful to regard the corresponding orthogonal webs. See Fig.~\ref{F13} for the web generated by $K$
and Fig.~\ref{F14} for the web generated by $\widetilde{K}$.

\begin{figure}[t]\centering
  \begin{minipage}[t]{.45\textwidth}
   \begin{center}
    \includegraphics[width=6.9cm]{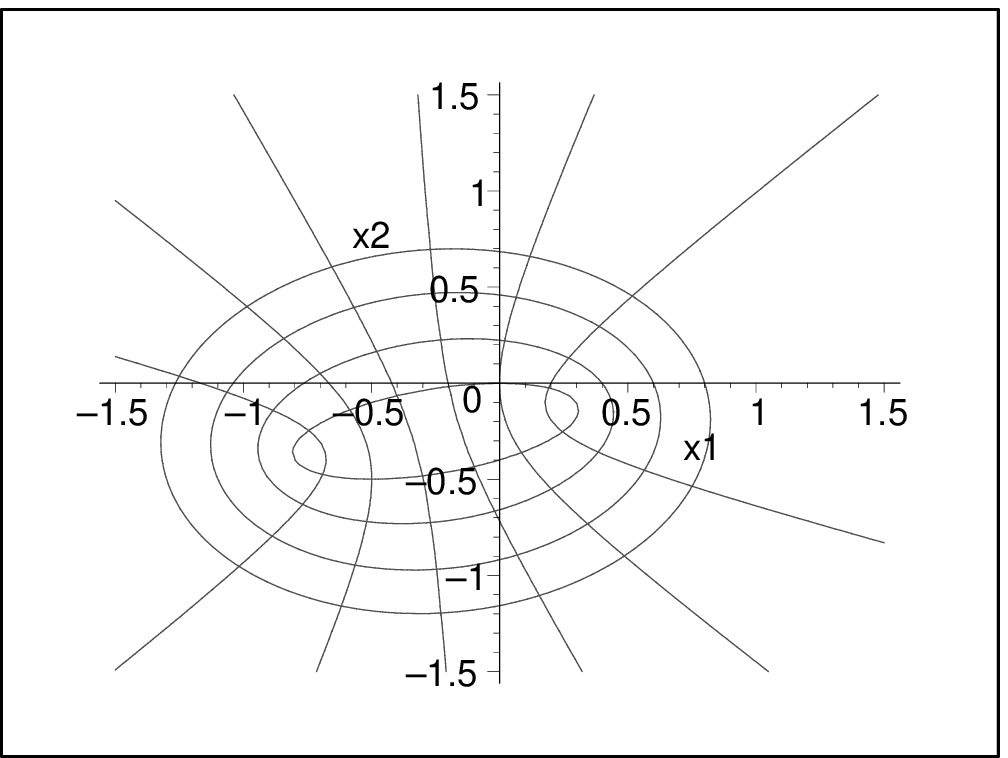}
     \caption[Elliptic-hyperbolic web 1]{Web corresponding to $K$.}
     \label{F13}
   \end{center}
  \end{minipage}
  \qquad
  \begin{minipage}[t]{.45\textwidth}
   \begin{center}
    \includegraphics[width=6.9cm]{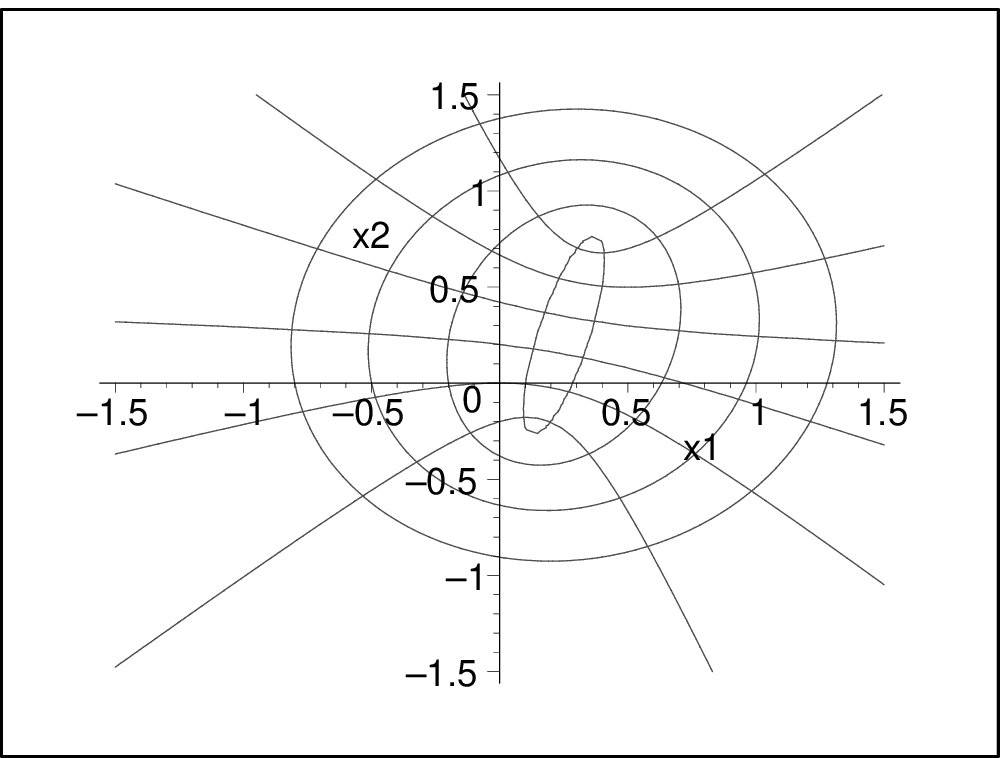}
    \caption[Elliptic-hyperbolic web 2]{Web corresponding to $\widetilde{K}$.}
    \label{F14}
   \end{center}
  \end{minipage}
\end{figure}

Consulting Table~\ref{T13} will reveal that $p_1 \in U_1$ and $p_2 \in U_3$. Table~\ref{T14} gives the appropriate
right moving frame for each point. Namely, for $p_1$ (the web in Fig.~\ref{F13})
\begin{gather}
 \theta = - \tfrac{1}{2}   \arctan \left (
\tfrac{1}{2} \right ) \ \mathrm{radians} \approx -13 \
\mathrm{degrees},\nonumber\\  a = \tfrac{1}{4} \cos \left
( \tfrac{1}{2} \arctan \left ( \tfrac{1}{2} \right ) \right ) +
\tfrac{1}{4} \sin \left ( \tfrac{1}{2} \arctan \left ( \tfrac{1}{2}
\right ) \right ) \approx 0.30,\nonumber\\  b =
\tfrac{1}{4} \cos \left ( \tfrac{1}{2} \arctan \left ( \tfrac{1}{2}
\right ) \right ) - \tfrac{1}{4} \sin \left ( \tfrac{1}{2} \arctan
\left ( \tfrac{1}{2} \right ) \right ) \approx 0.18.\label{TFs1}
\end{gather}
Substituting (\ref{TFs1}) into the action (\ref{ActionSE2R2}) gives the transformation
\begin{gather*}
 \bar{x}^1 = x^1 \cos \left ( \tfrac{1}{2} \arctan \left
( \tfrac{1}{2} \right ) \right ) + x^2 \sin \left ( \tfrac{1}{2}
\arctan \left ( \tfrac{1}{2} \right ) \right )\\
\phantom{\bar{x}^1 =}{} +  \tfrac{1}{4} \cos \left (
\tfrac{1}{2} \arctan \left ( \tfrac{1}{2} \right ) \right ) +
\tfrac{1}{4} \sin \left ( \tfrac{1}{2} \arctan \left ( \tfrac{1}{2}
\right ) \right ),\\
 \bar{x}^2 = - x^1 \sin
\left ( \tfrac{1}{2} \arctan \left ( \tfrac{1}{2} \right ) \right ) +
x^2 \cos \left ( \tfrac{1}{2} \arctan \left ( \tfrac{1}{2} \right )
\right )\\
\phantom{\bar{x}^2 =}{} + \tfrac{1}{4}
\cos \left ( \tfrac{1}{2} \arctan \left ( \tfrac{1}{2} \right ) \right
) - \tfrac{1}{4} \sin \left ( \tfrac{1}{2} \arctan \left ( \tfrac{1}{2}
\right ) \right ),
\end{gather*}
which maps the singular points (foci) in Fig.~\ref{F13} to the $x^1$-axis so that they are equally distant from the origin.
Similarly, consulting Table~\ref{T14} we get for $p_2$ (the web in Fig.~\ref{F14})
\begin{gather}
 \theta = - \tfrac{1}{2} \arctan (2) - \tfrac{\pi}{4} \
\mathrm{radians} \approx -77 \ \mathrm{degrees},\label{TFs2}\\
 a = - \tfrac{\sqrt{2}}{4} \cos \left ( \tfrac{1}{2}
\arctan(2) \right ) \approx -0.30,\qquad b =
\tfrac{\sqrt{2}}{4} \sin \left ( \tfrac{1}{2} \arctan(2) \right )
\approx 0.18.\nonumber
\end{gather}
Substituting (\ref{TFs2}) into the action (\ref{ActionSE2R2}) gives the transformation
\begin{gather*}
 \bar{x}^1 = \cos \left ( \frac{\arctan ( 2 )}{2}
\right ) \left ( \frac{x^1 + x^2}{\sqrt{2}} - \frac{1}{2 \sqrt{2}}
\right ) +  \sin \left ( \frac{\arctan ( 2 )}{2} \right ) \left (
\frac{-x^1 + x^2}{\sqrt{2}} \right ),\\
\bar{x}^2 = \cos \left ( \frac{\arctan ( 2 )}{2} \right ) \left (
\frac{- x^1 + x^2}{\sqrt{2}} \right ) +  \sin \left ( \frac{\arctan
( 2 )}{2} \right ) \left ( \frac{-x^1 - x^2}{\sqrt{2}} + \frac{1}{2
\sqrt{2}} \right ),
\end{gather*}
mapping the singular points (foci) in Fig.~\ref{F14} to the $x^1$-axis so that they are equally distant from the origin.

In order to determine the canonical form for both $p_1$ and $p_2$, simply appeal to (\ref{PPlus}) and (\ref{PMinus}).
The resulting canonical forms are
\begin{gather*}
 p_1 \equiv \left ( \frac{5 + \sqrt{5}}{4}, \frac{5 - \sqrt{5}}{4}, 0, 0, 0, 4 \right ) , \hspace{1.0cm} p_2 \equiv \left ( \frac{7 - \sqrt{5}}{4}, \frac{7 + \sqrt{5}}{4}, 0, 0, 0, -4 \right ),
\end{gather*}
which correspond to Killing tensors that generate elliptic-hyperbolic webs with singular values on the $x^1$-axis equally
distant from the origin in $\mathbb{E}^2$.
\end{example}

\subsection[The $\mathrm{SE}(3)$-equivalence of Killing two-tensors]{The $\boldsymbol{\mathrm{SE}(3)}$-equivalence of Killing two-tensors}

The components of a general Killing two-tensor def\/ined in Euclidean space are given by
\begin{gather}
K_{11} = A_{{1}}-2\,B_{{1,2}}x^3+2\,B_{{1,3}}x^2+C_{{2}}{(x^3)}^{2}+C_{{3}}{(x^2)}^{2}-2\, \gamma_{{1}}x^2x^3,\nonumber\\
K_{22} = A_{{2}}-2\,B_{{2,3}}x^1+2\,B_{{2,1}}x^3+C_{{3}}{(x^1)}^{2}+C_{{1}}{(x^3)}^{2}-2\, \gamma_{{2}}x^3 x^1,\nonumber\\
K_{33} = A_{{3}}-2\,B_{{3,1}}x^2+2\,B_{{3,2}}x^1+C_{{1}}{(x^2)}^{2}+C_{{2}}{(x^1)}^{2}-2\, \gamma_{{3}}x^1 x^2,\nonumber\\
K_{23} = \alpha_{{1}}+B_{{3,1}}x^3-B_{{2,1}}x^2+ \beta_1 x^1+ \left( \gamma_{{3}}x^3+\gamma_{{2}}x^2-\gamma_{{1}}x^1 \right) x^1 -C_{{1}}x^2x^3,\nonumber\\
K_{31} = \alpha_{{2}}+B_{{1,2}}x^1-B_{{3,2}}x^3+ \beta_2 x^2+ \left( \gamma_{{1}}x^1+\gamma_{{3}}x^3-\gamma_{{2}}x^2 \right) x^2 -C_{{2}}x^3x^1,\nonumber\\
K_{12} = \alpha_{{3}}+B_{{2,3}}x^2-B_{{1,3}}x^1 - \left( \beta_1 + \beta_2 \right) x^3+ \left( \gamma_{{2}}x^2+\gamma_{{1}}x^1-\gamma_{{3}}x^3 \right) x^3 -C_{{3}}x^1x^2.\label{KTE3}
\end{gather}
 The $\mathrm{SE}(3)$ equivalence problem on these Killing tensors concerns the action of a $6$-dimensional group acting on a $20$-dimensional vector space. As a result,
any explicit representation of the induced action on the twenty parameters def\/ining a Killing-two tensor will be
very large. To obtain such a representation, we must f\/irst consider the action of the proper Euclidean group for
$\mathbb{E}^3$, mapping a point $(x^1, x^2, x^3) \in \mathbb{E}^3$ to
\begin{gather}
\bar{x}^1 = \cos \left( \theta_{{{x}^{3}}} \right) \cos \left( \theta_{{{x}^{2}}}
 \right) x^1 +\cos \left( \theta_{{{x}^{3}}} \right) \sin \left( \theta_{
{{x}^{2}}} \right) \sin \left( \theta_{{x^1}} \right) {x}^{2}+\cos
 \left( \theta_{{{x}^{3}}} \right) \sin \left( \theta_{{{x}^{2}}}
 \right) \cos \left( \theta_{{x^1}} \right) {x}^{3}\nonumber\\
\phantom{\bar{x}^1 =}{} -\sin \left( \theta_{
{{x}^{3}}} \right) \cos \left( \theta_{{x^1}} \right) {x}^{2}+\sin
 \left( \theta_{{{x}^{3}}} \right) \sin \left( \theta_{{x^1}} \right) {x
}^{3}+a,\nonumber\\
\bar{x}^2 = \sin \left( \theta_{{{x}^{3}}} \right) \cos \left( \theta_{{{x}^{2}}}
 \right) x^1 +\sin \left( \theta_{{{x}^{3}}} \right) \sin \left( \theta_{
{{x}^{2}}} \right) \sin \left( \theta_{{x^1}} \right) {x}^{2}+\sin
 \left( \theta_{{{x}^{3}}} \right) \sin \left( \theta_{{{x}^{2}}}
 \right) \cos \left( \theta_{{x^1}} \right) {x}^{3}\nonumber\\
\phantom{\bar{x}^2 =}{}+\cos \left( \theta_{
{{x}^{3}}} \right) \cos \left( \theta_{{x^1}} \right) {x}^{2}-\cos
 \left( \theta_{{{x}^{3}}} \right) \sin \left( \theta_{{x^1}} \right) {x
}^{3}+b,\nonumber\\
\bar{x}^3 = -\sin \left( \theta_{{{x}^{2}}} \right) x+\cos \left( \theta_{{{x}^{2}
}} \right)  \left( \sin \left( \theta_{{x^1}} \right) {x}^{2}+\cos
 \left( \theta_{{x^1}} \right) {x}^{3} \right) +c.\label{ActionSE3R3}
\end{gather}
This particular representation is given by f\/irst rotating about the $x^3$-axis $(\theta_{x^3})$, followed by the
$x^2$-axis $(\theta_{x^2})$ and ending with the rotation about the $x^1$-axis $(\theta_{x^1})$. In this case, the group
parameters are given by $(\theta_{x^1},\theta_{x^2}, \theta_{x^3}, a, b, c)$.

To obtain the corresponding action on the parameters of $\mathcal{K}^2( \mathbb{E}^3 )$, we must determine
the ef\/fect of the push forward for (\ref{ActionSE3R3}) on the components $K^{ij}$ (\ref{KTE3}). Such an action
has been computed, however due to sheer size it cannot appear here. In order to illustrate, we have that
the parameter $B_{1,3}$ transforms like
\begin{gather*}
\overline{B_{1,3}}   =   C_{{1}}b \left( \cos \left( \theta_{{{x}^{2}}} \right)  \right) ^{2}-C
_{{2}} \left( \cos \left( \theta_{{{x}^{2}}} \right)  \right) ^{2}b-B_
{{2,1}}\sin \left( \theta_{{x^1}} \right) \cos \left( \theta_{{{x}^{3}}}
 \right) -B_{{3,1}}\cos \left( \theta_{{x^1}} \right) \cos \left( \theta
_{{{x}^{3}}} \right)\\
 \phantom{\overline{B_{1,3}}   =}{}   +B_{{3,2}}\cos \left( \theta_{{{x}^{2}}} \right)
\sin \left( \theta_{{{x}^{3}}} \right) -C_{{1}}\cos \left( \theta_{{{x
}^{2}}} \right) c\sin \left( \theta_{{{x}^{2}}} \right) \sin \left(
\theta_{{{x}^{3}}} \right)\\
\phantom{\overline{B_{1,3}}   =}{} -B_{{3,2}}\cos \left( \theta_{{{x}^{2}}}
 \right) \sin \left( \theta_{{{x}^{3}}} \right)  \left( \cos \left(
\theta_{{x^1}} \right)  \right) ^{2}+B_{{3,1}}\cos \left( \theta_{{x^1}}
 \right) \cos \left( \theta_{{{x}^{3}}} \right)  \left( \cos \left(
\theta_{{{x}^{2}}} \right)  \right) ^{2}\\
\phantom{\overline{B_{1,3}}   =}{} +B_{{2,1}}\sin \left( \theta_{
{x^1}} \right) \cos \left( \theta_{{{x}^{3}}} \right)  \left( \cos
 \left( \theta_{{{x}^{2}}} \right)  \right) ^{2}+C_{{2}}\cos \left(
\theta_{{{x}^{2}}} \right) \sin \left( \theta_{{{x}^{3}}} \right) \sin
 \left( \theta_{{{x}^{2}}} \right) c\\
\phantom{\overline{B_{1,3}}   =}{}    +2 \gamma_{{2}} \left( \cos
 \left( \theta_{{{x}^{2}}} \right)  \right) ^{2}\sin \left( \theta_{{{
x}^{3}}} \right) \cos \left( \theta_{{x^1}} \right) c+\gamma_{{2}}\sin
 \left( \theta_{{x^1}} \right) \cos \left( \theta_{{{x}^{3}}} \right)
\sin \left( \theta_{{{x}^{2}}} \right) c\\
\phantom{\overline{B_{1,3}}   =}{}  +2\gamma_{{1}}\sin \left(
\theta_{{x^1}} \right) \cos \left( \theta_{{{x}^{2}}} \right) c\cos
 \left( \theta_{{x^1}} \right) \sin \left( \theta_{{{x}^{2}}} \right)
\sin \left( \theta_{{{x}^{3}}} \right) \\
\phantom{\overline{B_{1,3}}   =}{} +2\gamma_{{1}} \left( \cos
 \left( \theta_{{x^1}} \right)  \right) ^{2}\cos \left( \theta_{{{x}^{3}
}} \right) \cos \left( \theta_{{{x}^{2}}} \right) c
    -C_{{3}}\sin
 \left( \theta_{{x^1}} \right) \cos \left( \theta_{{{x}^{2}}} \right) c
\cos \left( \theta_{{x^1}} \right) \cos \left( \theta_{{{x}^{3}}}
 \right) \\
 \phantom{\overline{B_{1,3}}   =}{} -\gamma_{{3}}\sin \left( \theta_{{{x}^{2}}} \right) c\cos
 \left( \theta_{{x^1}} \right) \cos \left( \theta_{{{x}^{3}}} \right)
     -C
_{{2}} \left( \cos \left( \theta_{{x^1}} \right)  \right) ^{2}\cos
 \left( \theta_{{{x}^{2}}} \right) c\sin \left( \theta_{{{x}^{2}}}
 \right) \sin \left( \theta_{{{x}^{3}}} \right)\\
\phantom{\overline{B_{1,3}}   =}{} +B_{{2,3}}\cos \left(
\theta_{{{x}^{2}}} \right) \cos \left( \theta_{{x^1}} \right) \sin
 \left( \theta_{{{x}^{2}}} \right) \cos \left( \theta_{{{x}^{3}}}
 \right) \sin \left( \theta_{{x^1}} \right)\\
\phantom{\overline{B_{1,3}}   =}{} +C_{{3}}\cos \left( \theta_{
{{x}^{2}}} \right) \sin \left( \theta_{{{x}^{3}}} \right) \sin \left(
\theta_{{{x}^{2}}} \right) c \left( \cos \left( \theta_{{x^1}} \right)
 \right) ^{2}+2\,\gamma_{{3}} \left( \cos \left( \theta_{{{x}^{2}}}
 \right)  \right) ^{2}\sin \left( \theta_{{{x}^{3}}} \right) \sin
 \left( \theta_{{x^1}} \right) c\\
\phantom{\overline{B_{1,3}}   =}{} +C_{{2}}\sin \left( \theta_{{x^1}}
 \right) \cos \left( \theta_{{{x}^{2}}} \right) c\cos \left( \theta_{{
x^1}} \right) \cos \left( \theta_{{{x}^{3}}} \right) + \left( \cos
 \left( \theta_{{{x}^{2}}} \right)  \right) ^{2}\cos \left( \theta_{{{
x}^{3}}} \right) B_{{1,3}}\cos \left( \theta_{{x^1}} \right)\\
\phantom{\overline{B_{1,3}}   =}{} +B_{{3,2}}
\cos \left( \theta_{{{x}^{2}}} \right) \cos \left( \theta_{{x^1}}
 \right) \sin \left( \theta_{{{x}^{2}}} \right) \cos \left( \theta_{{{
x}^{3}}} \right) \sin \left( \theta_{{x^1}} \right) \\
\phantom{\overline{B_{1,3}}   =}{} -\beta_{{1}}\cos
 \left( \theta_{{{x}^{2}}} \right) \sin \left( \theta_{{{x}^{2}}}
 \right) \cos \left( \theta_{{{x}^{3}}} \right)  \left( \cos \left(
\theta_{{x^1}} \right)  \right) ^{2}
    -\beta_{{1}}\cos \left( \theta_{{{x}
^{2}}} \right) \sin \left( \theta_{{{x}^{3}}} \right) \sin \left(
\theta_{{x^1}} \right) \cos \left( \theta_{{x^1}} \right) \\
\phantom{\overline{B_{1,3}}   =}{} -2 \left( \cos
 \left( \theta_{{{x}^{2}}} \right)  \right) ^{2}\gamma_{{1}}b\sin
 \left( \theta_{{x^1}} \right) \cos \left( \theta_{{x^1}} \right)
    +2
\gamma_{{3}}\cos \left( \theta_{{{x}^{2}}} \right) b\sin \left( \theta
_{{x^1}} \right) \sin \left( \theta_{{{x}^{2}}} \right) \\
\phantom{\overline{B_{1,3}}   =}{} +2\cos \left(
\theta_{{{x}^{2}}} \right) \gamma_{{2}}b\sin \left( \theta_{{{x}^{2}}}
 \right) \cos \left( \theta_{{x^1}} \right)
    -C_{{3}} \left( \cos \left(
\theta_{{{x}^{2}}} \right)  \right) ^{2}b \left( \cos \left( \theta_{{
x^1}} \right)  \right) ^{2}\\
\phantom{\overline{B_{1,3}}   =}{} -\gamma_{{2}}\sin \left( \theta_{{{x}^{3}}}
 \right) c\cos \left( \theta_{{x^1}} \right) -\gamma_{{3}}c\sin \left(
\theta_{{x^1}} \right) \sin \left( \theta_{{{x}^{3}}} \right)
    -B_{{3,1}}
\sin \left( \theta_{{x^1}} \right) \sin \left( \theta_{{{x}^{2}}}
 \right) \sin \left( \theta_{{{x}^{3}}} \right) \\
 \phantom{\overline{B_{1,3}}   =}{} -\cos \left( \theta_{{
{x}^{2}}} \right) \gamma_{{1}}c\cos \left( \theta_{{{x}^{3}}} \right)
+\cos \left( \theta_{{{x}^{2}}} \right) \cos \left( \theta_{{{x}^{3}}}
 \right) \beta_{{2}}\sin \left( \theta_{{{x}^{2}}} \right)\\
\phantom{\overline{B_{1,3}}   =}{} +\beta_{{1}
}\cos \left( \theta_{{{x}^{2}}} \right) \sin \left( \theta_{{{x}^{2}}}
 \right) \cos \left( \theta_{{{x}^{3}}} \right) -B_{{2,3}}\cos \left(
\theta_{{{x}^{2}}} \right) \sin \left( \theta_{{{x}^{3}}} \right)
 \left( \cos \left( \theta_{{x^1}} \right)  \right) ^{2}\\
\phantom{\overline{B_{1,3}}   =}{} +B_{{2,1}}\cos
 \left( \theta_{{x^1}} \right) \sin \left( \theta_{{{x}^{2}}} \right)
\sin \left( \theta_{{{x}^{3}}} \right) + \left( \cos \left( \theta_{{{
x}^{2}}} \right)  \right) ^{2}\cos \left( \theta_{{{x}^{3}}} \right) B
_{{1,2}}\sin \left( \theta_{{x^1}} \right)\\
\phantom{\overline{B_{1,3}}   =}{} +C_{{2}} \left( \cos \left(
\theta_{{{x}^{2}}} \right)  \right) ^{2}b \left( \cos \left( \theta_{{
x^1}} \right)  \right) ^{2}-C_{{1}}b
\end{gather*}
under this action. In addition to this, there are $19$ more larger formulas that determine such an action. As a result,
computing the moving frame associated with a cross-section proves to be a challenge. To this end, such a moving frame
has not been constructed. In many cases, even upon restricting one's attention
to particular invariant subspaces, it remains a very dif\/f\/icult problem, however some success has resulted in this
context.

Due to the sheer size of the system, it may seem prudent to seek an alternative method to solve the corresponding
equivalence problem. In doing so, invariant functions for this action have been calculated without resorting to the action.
In particular, the invariant condition on the inf\/initesimal generators leads to a system of f\/irst order linear
homogeneous PDEs. The solution to these PDEs yields the invariant functions. Utilizing this technique, a complete set
of invariant functions has been computed and used to distinguish between the type of orthogonal web generated by a given
valence-two Killing tensor in Euclidean space. Consult \cite{HMS 2004} for the details concerning the calculation
and application of the invariant functions.

The problem of distinguishing between the orbits is of a more subtle nature then distinguishing between the type of web,
and may be very dif\/f\/icult without resorting to the action. For example, if we choose a cross-section and see that the
invariant functions take a unique value at each point on the cross-section, then we know that the invariant functions
will distinguish between each orbit through the cross-section. The dif\/f\/iculty lies in determining whether a particular
point belongs to an orbit that intersects the given cross-section. For this, it may be necessary to resort to the action.

\section[Hamilton-Jacobi theory]{Hamilton--Jacobi theory} \label{HJTs}

 In this section we shall brief\/ly review the underlying idea of the Hamilton--Jacobi theory of orthogonal
 separation of variables and establish the requisite language to be used in what follows. Let $(M, {\bf g})$ be
an $n$-dimensional  pseudo-Riemannian manifold of constant curvature. Recall that a Hamiltonian
system def\/ined by a natural Hamiltonian function with a scalar potential $V$, which can be written as
\begin{gather}
 \label{H}
 H ({\bf q},{\bf p}) = \frac{1}{2}g^{ij}({\bf q})p_ip_j + V({\bf q}),
\end{gather}
can in many cases  be  integrated by quadratures by considering the corresponding
Hamilton--Jacobi equation (HJE). Here $g^{ij}$ are the contravariant components of the corresponding metric tensor $\bf g$ and
$({\bf q},{\bf p}) \in T^*M$ are the canonical position-momenta coordinates. The
procedure consists of  a canonical coordinate transformation (CT) $T : ({\bf q},{\bf p})$
$\rightarrow$ $({\bf u}, {\bf v})$ to  {\em separable coordinates} (SC) $({\bf u}, {\bf v})$, with respect
to which the Hamilton--Jacobi equation
\begin{gather}
\frac{1}{2} g^{ij}({\bf u}) \frac{\partial W}{\partial u^i} \frac{\partial W}{\partial u^j} + V({\bf u}) = E, \qquad v_j = \frac{\partial W}{\partial u^i},
\label{HJE}
\end{gather}
admits a {\it complete integral} (CI) $W ({\bf u},{\bf c})$, satisfying the
non-degeneracy condition:
\[
\det\|\partial^2 W /\partial u^i\partial
c_j\|_{n\times n} \not= 0,
\] where ${\bf c} = (c_1,\ldots, c_n)$ is  a constant
vector. The function $W$ is usually sought in the form
\[
W({\bf u}, {\bf c}) =
\sum_{i=1}^nW_i(u^i, {\bf c}),
\]  which is the essence of the {\it additive
separation ansatz}. In view of Jacobi's theorem, once $W$ has been found, the
integral curves of the f\/low generated by (\ref{H}) can be determined  from the
equations
\[
v_i = \frac{\partial W}{\partial u^i}, \qquad b_j = \frac{\partial
W}{\partial c_j}, \qquad t - t_0 = \frac{\partial W}{\partial E},
\]
 where $i =
1, \ldots, n, $  $j = 1, \ldots, n-1.$ The inverse canonical transformation
$({\bf u},{\bf v})$ $\rightarrow$ $({\bf q},{\bf p})$ yields the solution in
terms of the original position-momenta coordinates $({\bf q},{\bf p})$.
Geometrically, the Hamilton--Jacobi equation and its solution $W$ can be
interpreted as follows (see Benenti \cite{Be1}): In a neighborhood of a regular
point the equation (\ref{HJE}) determines a hypersurface ${\cal H}
\subset T^*M$, while the set of equations $v_i = \partial W/\partial u_i$
 determine  a Lagrangian submanifold $\Lambda \subset T^*M$, as an image of a closed one-form
 $dW$. Therefore $W$ is a solution to (\ref{HJE}) if\/f ${\cal H} \subset \Lambda$.
The Hamilton--Jacobi theory of {\em orthogonal} separation of variables is based on point transformations to
separable coordinates, namely the transformations of the form
$u^i = u^i({\bf q}), \, i = 1,\ldots, n.$  Moreover, the point transformation in this context is
{\em  (non-)orthogonal} if\/f the metric tensor $\bf g$ of (\ref{H}) is (non-) diagonal with respect
to the separable coordinates $u^1, \ldots, u^n$.

The existence of orthogonal  separable coordinates $({\bf u},{\bf v})$ is usually guaranteed by
the existence and geometric properties of Killing tensors associated with the system def\/ined by (\ref{H}).

The following criterion due to Benenti \cite{Be1} generalizes the famous theorem proved by Eisenhart for geodesic
Hamiltonians \cite{Ei34}:
 \begin{theorem}
 \label{t1}
 The Hamiltonian system defined by $(\ref{H})$ is orthogonally separable if and only if there exists a valence two Killing tensor
$\bf K$ with pointwise simple and real eigenvalues, orthogonally integrable eigenvectors such that
\begin{gather}
\label{CC}
\mbox{d}({\bf \hat{K}}
\mbox{d}V) = 0,
\end{gather}
where the linear operator $\bf \hat{K}$ is given by ${\bf \hat{K}}:= {\bf K}{\bf g}$.
 \end{theorem}
 The hypothesis of Theorem \ref{t1} implies that the Hamiltonian system in question admits a f\/irst integral quadratic in
 the momenta of the form:
 \begin{gather*}
 F({\bf q}, {\bf p}) = \tfrac{1}{2}K^{ij}({\bf q})p^ip^j + U({\bf q}),
 \end{gather*}
 where $K^{ij}$ are the components of the Killing tensor $\bf K$ and $\mbox{d}U = \hat{K}\mbox{d}V$, while $\hat{K}$ is the $(1,1)$-tensor obtained from $\bf K$ by lowering one index. As functions of the position coordinates ${\bf q} = (q^1,\ldots, q^n)$ the components $K^{ij}$, $i,j=1,\ldots, n$ satisfy the {\em Killing tensor equation}:
 \begin{gather*}
    [{\bf g}, {\bf K}]^{ijk} = g^{(ij}_{\,\;\;\;,\ell}K^{k)\ell} - K^{(ij}_{\;\;\;\,,\ell}g^{k)\ell} = 0,
 \end{gather*}
where $[~,~]$ is the Schouten bracket.
It must be mentioned that in the case when the under\-lying manifold is of dimension two, the condition of orthogonal
integrability of eighenvectors in Theorem \ref{t1} can be dropped (it satisf\/ies them automatically). On the other hand,
the condition that the eigenvalues of ${\bf K}$ be real is essential, since they can be complex
in general when the underlying manifold is pseudo-Riemannian. Note that Theorem~\ref{t1} is the key result that allows us
to connect naturally  the Hamilton--Jacobi theory of orthogonal separation of variables with the study of vector spaces of
Killing tensors under the action of the corresponding isometry groups.
\begin{example}  Let $(M, {\bf g}) = \mathbb{E}^2$. Then the components of the  general solution (with respect to
Cartesian coordinates ${\bf x} = (x^1, x^2)$
to the Killing tensor equation in this case can be expressed as follows:
\begin{gather}
K^{11}({\bf x})  =  \alpha_1 + 2 \alpha_4 x^2 + \alpha_6 (x^2)^2, \qquad
K^{12}({\bf x}) =K^{21}({\bf x})  =   \alpha_3 - \alpha_4 x^1 - \alpha_5 x^2 - \alpha_6 x^1 x^2, \nonumber\\
K^{22}({\bf x})  = \alpha_2 + 2 \alpha_5 x^1 + \alpha_6 (x^1)^2.\label{e1}
\end{gather}
\end{example}

The solution space to the Killing tensor equation given by (\ref{e1}) in this case is
nothing but  the vector space of Killing two tensors def\/ined in ${\mathbb E}^2$ which we denote here by
${\cal K}^2(\mathbb{E}^2)$. The orthogonal coordinate systems are
then def\/ined by the $n$ foliations, the leaves of which are
$(n-1)$-dimensional hypersurfaces orthogonal to the eigenvectors of $\bf K$. Such geometric structures def\/ined by
the Killing tensors having the properties prescribed in Theorem \ref{t1} are called {\em orthogonal coordinate webs}.

Thus, employing the Hamilton--Jacobi theory of orthogonal separation of variables to solve  a Hamiltonian system def\/ined by a Hamiltonian whose potential satisf\/ies the hypothesis of Theorem \ref{t1} boils down to solving the following two fundamental problems:
\begin{enumerate}\itemsep=0pt

\item Classify each Killing tensor $\bf K$  with distinct eigenvalues and orthogonally
integrable eigenvectors, which is  compatible with a given potential $V$ via the compatibility condition (\ref{CC}).
In this context the classif\/ication problem is equivalent to the problem of the determination of the orthogonal coordinate webs generated by $\bf K$. The most natural framework of solving this problem is via considering the orbit problem of the corresponding isometry group
 acting in the vector space of Killing tensors of valence two.

\item Once the f\/irst problem is solved, one next has to determine the transformation(s) of the corresponding orthogonal coordinate webs to their canonical forms.
\end{enumerate}

\begin{example} \label{exya} Consider the 2nd integrable case of Yatsun def\/ined by the natural Hamiltonian with the potential $V$ given by:
\begin{gather}
V (q^1,q^2) =  -2 \left((q^1)^4 + 2 (q^1)^2 (q^2)^2 +
\frac{2\lambda}{g^2}(q^2)^4\right)  \nonumber\\
\phantom{V (q^1,q^2) =}{}+ 4 \left((q^1)^3 + q^1
(q^2)^2\right) - 2\left((q^1)^2 + (q^2)^2\right). \label{YP2}
\end{gather}
 It has
been shown (see \cite{MST 2002JMP, MST04} for the references)  that the Hamiltonian system is completely integrable if $g^2= 2\lambda$ admitting the following additional
f\/irst integral quadratic in the momenta:
 \begin{gather*}
 F_2  =  \left((q^2)^2 + \frac{3}{4}\right)p_1^2 - (2q^1 -1)q^2p_1p_2 +
(q^1-1)q^1p_2^2 \nonumber\\
\phantom{F_2  =}{} -3(q^1)^4 -
2(q^1)^2(q^2)^2 + (q^2)^4 + 6(q^1)^3 + 2(q^1)(q^2)^2 - 3(q^1)^2. 
 \end{gather*}
Taking into the account the formula (\ref{e1}), we conclude therefore that the Killing tensor $\bf K$ compatible with the potential given by (\ref{YP2}) is given by
\begin{gather}
K^{11}({\bf q})  =  \tfrac 34  + (q^2)^2, \qquad
K^{12}({\bf q}) =K^{21}({\bf q})  =    \tfrac 12 q^2 -  q^1 q^2, \nonumber\\
K^{22}({\bf q})  =   -q^1 + (q^1)^2.\label{e2}
\end{gather}
Now, in view of the above, in order  to solve the problem of orthogonal separability of the corresponding Hamilton--Jacobi equation and thus f\/ind exact solutions to the original Hamiltonian system, one has to determine 1) the type of orthogonal coordinate web that the Killing tensor given by (\ref{e2}) generates, 2) determine the transformation of the coordinates $(q^1,q^2)$ that put the Killing tensor (\ref{e2}) into its canonical form.
\end{example}

In what follows we shall show that the problems above can be solved in general within the context of Section~\ref{class}.
More specif\/ically, the f\/irst problem is essentially the equivalence problem, while the second problem is the canonical forms problem and the problem of the determination of the corresponding moving frame map(s) for a given cross-section.

\section{Fusion}\label{fusion}

In view of the material presented in Section~\ref{class}, it is evident now that the two basic problems of the Hamilton--Jacobi theory of orthogonal separation of variables presented in Section~\ref{HJTs} can be translated into the geometric language of group actions. Indeed, let ${\bf K} \in {\cal K}^2(M)$ be a Killing two-tensor  def\/ined on a pseudo-Riemannian manifold $(M,{\bf g})$ of constant curvature satisfying the hypothesis of Theorem~\ref{t1} and  $G$ -- the corresponding isometry group. Then the problem of solving the Hamiltonian system in question via orthogonal separation of variables (refer to Theorem~\ref{t1}) in the associated Hamilton--Jacobi equation reduces to following two problems:
\begin{enumerate}\itemsep=0pt
\item Determine the orbit of the group action $G \circlearrowright {\cal K}^2(M)$ that the Killing tensor ${\bf K}$ corresponds~to.

\item For a given cross-section determine the moving frame map that maps the point on the orbit corresponding to ${\bf K}$ to the intersection of the cross-section with the orbit (canonical form).
\end{enumerate}

To conf\/irm our claim with a proper illustration we now revisit Example~\ref{exya}.  Recall that the Killing tensor $\textbf{K}$,
see (\ref{e2}), compatible with the potential given by (\ref{YP2}) may be represented by the symmetric matrix
\begin{gather}
\left(
  \begin{array}{cc}
    \displaystyle \tfrac{3}{4} + (q^2)^2 & \displaystyle \tfrac{1}{2} q^2 - q^1 q^2 \vspace{2mm}\\
    \displaystyle \tfrac{1}{2} q^2 - q^1 q^2 & - q^1 + (q^1)^2
  \end{array}
\right). \label{YAPMAT}
\end{gather}
Comparing (\ref{YAPMAT}) with the general Killing tensor (\ref{MatrixKT}) will show that $\textbf{K}$ corresponds to the
point
\begin{gather*}
( \alpha_1, \alpha_2, \alpha_3, \alpha_4, \alpha_5, \alpha_6 ) = ( 3/4, 0, 0, 0, -1/2, 1 )
\end{gather*}
in the parameter space $\Sigma$. Appealing to Table~\ref{T7}, since
\begin{gather*}
\Delta_1 = (\alpha_6(\alpha_1 - \alpha_2) - \alpha_4^2 + \alpha_5^2)^2 + 4(\alpha_6 \alpha_3 + \alpha_4 \alpha_5)^2 = 1 \neq 0,
\end{gather*}
we see that the Killing tensor $\textbf{K}$ belongs to the invariant submanifold $E_3$, see (\ref{Mans}), i.e.\ a~generic 3-dimensional
orbit. Moreover, $\Delta_2 = \alpha_6 \neq 0$, and as a result $\textbf{K}$ belongs to the invariant submanifold
$E_3^2$, see (\ref{EHPA}), and so is an elliptic-hyperbolic web.

Consulting Table~\ref{T13}, it is clear that the point represented by $\textbf{K}$ lies in the chart
$( U_1, \varphi )$, since
\begin{gather*}
\iota_1 = \alpha_6 (\alpha_1 - \alpha_2) - \alpha_4^2 + \alpha_5^2 = 1 > 0.
\end{gather*}
Table \ref{T14} thus indicates that the associated moving frame map is given by
\begin{gather*}
\psi_1(\alpha_1, \ldots, \alpha_6) = \left(%
\begin{array}{c}
  \theta_1 \\
  a_1 \\
  b_1 \\
\end{array}%
\right) = \left(%
\begin{array}{c}
  \displaystyle - \frac{1}{2}  \arctan  \left ( \frac{2( \alpha_3 \alpha_6 + \alpha_4 \alpha_5 )}{ \alpha_6(\alpha_1 - \alpha_2) - \alpha_4^2 + \alpha_5^2 } \right ) \vspace{1mm}\\
  \displaystyle \frac{\alpha_5   \cos   \theta_1 - \alpha_4   \sin   \theta_1}{\alpha_6} \vspace{1mm}\\
  \displaystyle \frac{\alpha_4   \cos   \theta_1 + \alpha_5   \sin   \theta_1}{\alpha_6}
\end{array}
\right),
\end{gather*}
from (\ref{MF2EH}). Immediately, by substituting the parameters into the moving frame map, we get that the separable
coordinates for the system are shifted (along the $x^1$-axis) elliptic-hyperbolic coordinates
\begin{gather}
\label{gpars}
\theta_1 = 0, \qquad a_1 = -\tfrac{1}{2}, \qquad b_1 = 0,
\end{gather}
as is well known. Indeed, substituting (\ref{gpars}) into the action (\ref{ActionSE2R2}) on $\mathbb{E}^2$, gives the
transformation
\begin{gather}
\label{KTE3a}
\overline{q}^1 = q^1 - \tfrac 12,\qquad
\overline{q}^2 = q^2,
\end{gather}
that takes $\textbf{K}$ to its canonical form.

Finally, we may utilize (\ref{PPlus}), (\ref{PMinus}) and the invariant functions (\ref{MF2EH}) to get the explicit
formula for the canonical form, namely
\begin{gather*}
\textbf{K} \equiv \left(
  \begin{array}{cc}
    \tfrac{3}{4} + (\overline{q}^2)^2 & - \overline{q}^1 \overline{q}^2 \vspace{1mm}\\
    - \overline{q}^1 \overline{q}^2 & - \tfrac{1}{4} + (\overline{q}^1)^2 \\
  \end{array}
\right). \label{CanYAP}
\end{gather*}
The problem of integrating the Hamiltonian system def\/ined by (\ref{YP2}) via orthogonal separation of variables in the
associated Hamilton--Jacobi equation, can now be solved with the aid of the separable coordinates $(\overline{q}^1,\overline{q}^2)$
given by (\ref{KTE3a}). Indeed, it is easy to verify, for example, that the potential $V$ (\ref{YP2}) after the transformation to separable coordinates given by (\ref{KTE3a}) will satisfy  the Levi-Civita criterion.

\section{Conclusion}\label{conclusion}

In this paper we have demonstrated that the underlying ideas of the Hamilton--Jacobi theory of orthogonal separation of variables can be naturally described in the language of the geometric theory of group actions as applied to the study  of Killing tensors def\/ined in pseudo-Riemannian spaces of constant curvature. Although we have mainly used the Euclidean space as the underlying space for our studies, it is clear that the approach will work for other homogeneous spaces as underlying spaces where the vector spaces of Killing two-tensors are def\/ined. For example, the natural projection $\pi:\,SO(3) \rightarrow SO(3)/SO(2)$ gives rise to  the corresponding problem based
on the geometric study of ${\cal K}^2(\mathbb{S}^2)$, the vector space of Killing two-tensors def\/ined on two-sphere $\mathbb{S}^2 = SO(3)/SO(2)$. The application of the moving frames method to the orbit problem $SO(3) \circlearrowright {\cal K}^2(\mathbb{S}^2)$ will provide the mathematical background to study the Hamilton--Jacobi orthogonal separation of variables for the Hamiltonian systems def\/ined on the curved space $\mathbb{S}^2$.

\subsection*{Acknowledgements}

The research was supported in part by  National Sciences and Engineering Research
Council of Canada Discovery Grants.
The authors wish to thank the anonymous referees for useful comments.

\pdfbookmark[1]{References}{ref}

\LastPageEnding

\end{document}